\pdfoutput=1
\RequirePackage{ifpdf}
\ifpdf
\documentclass[11pt,pdftex]{article}
\else
\documentclass[11pt,dvips]{article}
\fi

 \usepackage{epsfig}
 \usepackage{color,graphics}
 \usepackage{latexsym,amsmath,amsfonts,amsthm,amssymb}
 \usepackage{euscript,amsbsy}
 \usepackage{graphicx}
 \usepackage{caption}

\usepackage{latexsym}
\usepackage{amscd}
\usepackage{color,graphics}
\usepackage{graphicx}

\usepackage{cite} 

\usepackage{ifpdf}   
\DeclareGraphicsExtensions{.pdf,.png,.jpg,.mps}

\newcommand{\bea}{\begin{eqnarray}}
\newcommand{\eea}{\end{eqnarray}}
\newcommand{\be}{\begin{equation}}
\newcommand{\ee}{\end{equation}}

\newcommand{\bb}[1]{{\mathbb{#1}}}

\newtheorem{theo}{Theorem}[section]
\newtheorem{prop}[theo]{Proposition}
\newtheorem{lem}[theo]{Lemma}
\newtheorem{cor}[theo]{Corollary}

\newtheorem{defi}[theo]{Definition}
\newtheorem{rem}[theo]{Remark}
\newtheorem{ass}[theo]{Assumption}

\newcommand{\bl}{\begin{lem}}
\newcommand{\el}{\end{lem}}
\newcommand{\bp}{\begin{prop}}
\newcommand{\ep}{\end{prop}}
\newcommand{\bt}{\begin{theo}}
\newcommand{\et}{\end{theo}}
\newcommand{\bpr}{\begin{proof}}
\newcommand{\epr}{\end{proof}}

\newcommand{\bc}{\begin{cor}}
\newcommand{\ec}{\end{cor}}

\newcommand{\br}{\begin{rem}}
\newcommand{\er}{\end{rem}}
\newcommand{\bd}{\begin{defi}}
\newcommand{\ed}{\end{defi}}
\newcommand{\bass}{\begin{ass}}
\newcommand{\eass}{\end{ass}}

\newlength{\pecettawidth}
\begin{document}
\title{Effects of communication efficiency and exit capacity on 
fundamental diagrams for pedestrian motion in an obscure tunnel -- a particle system approach}
\author{Emilio N.M.\ Cirillo$^{1}$, Matteo Colangeli$^{2}$, Adrian Muntean$^{3}$ \\\\
\footnotesize{$^1$Dipartimento di Scienze di Base e Applicate per
             l'Ingegneria,}\\ \footnotesize{Sapienza Universit\`a di Roma,
             Via A.\ Scarpa 16, 00161, Rome, Italy.}\\
\footnotesize{$^2$Gran Sasso Science Institute, Via. F. \ Crispi 7, 00167 L' Aquila, Italy}\\
\footnotesize{$^3$Department of Mathematics and Computer Science,}\\
             \footnotesize{(CASA) Centre for Analysis, Scientific computing and Applications,}\\
             \footnotesize{Institute for Complex Molecular Systems (ICMS)
             Eindhoven University of Technology,}\\
             \footnotesize{P.O.\ Box 513, 5600, MB Eindhoven, The Netherlands}\\
\small{emails: {emilio.cirillo@uniroma1.it};~{matteo.colangeli@gssi.infn.it};~{a.muntean@tue.nl}}\\}

\maketitle

\begin{abstract}
Fundamental diagrams describing the relation between pedestrians speed 
and density are key points in understanding pedestrian dynamics. 
Experimental data evidence the onset of complex behaviors in which the 
velocity decreases with the density and different logistic regimes are
identified. This paper addresses the issue of pedestrians transport and of fundamental diagrams  for a scenario involving the motion of pedestrians 
escaping from an obscure tunnel. 
We  capture the effects of the communication efficiency and 
the exit capacity by means of two thresholds controlling the rate 
at which particles (walkers, pedestrians) move on the lattice. 
Using a particle system model, we show that in absence of limitation in communication among 
pedestrians we reproduce
with good accuracy the standard fundamental diagrams, whose 
basic behaviors can be interpreted in terms of the exit capacity 
limitation. 
When the effect of a limited communication ability is considered, then  
interesting non--intuitive phenomena occur. Particularly, we shed light on
the loss of monotonicity of the typical speed--density curves, 
revealing the existence of a
pedestrians density optimizing the escape. 

We study both the discrete particle dynamics as well as the corresponding hydrodynamic limit (a porous medium equation and a transport (continuity) equation). We also point out the dependence of the effective transport coefficients on the two thresholds -- the essence of the microstructure information. 
\end{abstract}




\vfill


\section{Introduction}
\label{s:int}
\par\noindent
\emph{Fundamental diagrams} representing the dependence of the 
pedestrian speed on the local density are one of the basic methods in 
studying pedestrians dynamics. They contain macroscopic information useful to identify 
the key effects affecting the general behavior of pedestrian flows and to 
test the validity of pedestrian models. 
In \cite{FD,Weid}, e.g., their main properties are discussed and 
experimental tests are performed. In particular it is seen that 
many different effects, such as passing manoeuvres, space reduction, 
and internal friction, have to be taken into account to 
explain the main features of the diagrams. 

In this paper, we use \emph{Zero Range Processes} (ZRP),
originally proposed by Spitzer \cite{S},
to recover the same behaviors of the fundamental diagrams, 
excepting perhaps  the existence of an upper density above which the 
pedestrian velocity drops to zero. Our attention focuses on pedestrians moving in dark corridors, where the lack of visibility hinders them to find the exit. This research line follows a similar path as in \cite{Bell1,Bell2,Pareschi,Albi}, where the authors used a kinetic formulation to investigate the role of the leaders to control crowds evacuation when visibility is reduced, and extends our previous works on this topic;  compare e.g. \cite{CM01,CM02,CM03} (group formation and cooperation in the dark).

For the current  framework,  we assume that 
more particles can occupy the same site of a one--dimensional array of discrete positions (modeling a long dark corridor) and 
no interaction among the 
individuals takes place. The dynamics of the system is determined 
only by the \emph{escape rate}, namely, the frequency at which 
a site releases the individuals. 
The key idea in our model is to assume that the escape rate is proportional 
to the number of individuals on the site up to an \emph{saturation
threshold} above which such a rate stays constant. The second 
ingredient we use is that escape is maintained low until a certain 
\emph{activation threshold} is reached. 

\begin{figure}
\begin{picture}(400,80)(-70,0)
\setlength{\unitlength}{.026cm}
\thicklines
\put(10,20){\line(1,0){120}}
\put(10,50){\line(1,0){120}}
\thinlines
\put(55,20){\line(0,1){30}}
\put(85,20){\line(0,1){30}}
\put(82.5,32.5){\circle*{5}}
\put(82.5,32.5){\circle{15}}
\put(67.5,40){\circle*{5}}
\put(67.5,40){\circle{15}}
\put(65.5,27.5){\circle*{5}}
\put(65.5,27.5){\circle{15}}
\thicklines
\put(10,70){\line(1,0){120}}
\put(10,100){\line(1,0){120}}
\thinlines
\put(55,70){\line(0,1){30}}
\put(85,70){\line(0,1){30}}
\put(82.5,82.5){\circle*{5}}
\put(82.5,82.5){\circle{15}}
\put(67.5,90){\circle*{5}}
\put(67.5,90){\circle{15}}
\put(65.5,77.5){\circle*{5}}
\put(65.5,77.5){\circle{15}}
\put(72.5,82.5){\circle*{5}}
\put(72.5,82.5){\circle{15}}
\thicklines
\put(210,20){\line(1,0){120}}
\put(210,50){\line(1,0){120}}
\thinlines
\put(255,20){\line(0,1){30}}
\put(285,20){\line(0,1){30}}
\put(258,27.5){\circle*{5}}
\put(258,32.5){\circle*{5}}
\put(258,47.5){\circle*{5}}
\put(282.5,27.5){\circle*{5}}
\put(282.5,37.5){\circle*{5}}
\put(282.5,42.5){\circle*{5}}
\put(282.5,47.5){\circle*{5}}
\thicklines
\put(210,70){\line(1,0){120}}
\put(210,100){\line(1,0){120}}
\thinlines
\put(255,70){\line(0,1){30}}
\put(285,70){\line(0,1){30}}
\put(258,77.5){\circle*{5}}
\put(258,72.5){\circle*{5}}
\put(258,82.5){\circle*{5}}
\put(258,87.5){\circle*{5}}
\put(258,92.5){\circle*{5}}
\put(258,97.5){\circle*{5}}
\put(282.5,77.5){\circle*{5}}
\put(282.5,72.5){\circle*{5}}
\put(282.5,82.5){\circle*{5}}
\put(282.5,87.5){\circle*{5}}
\put(282.5,92.5){\circle*{5}}
\put(282.5,97.5){\circle*{5}}
\put(277,72.5){\circle*{5}}
\put(272.5,87.5){\circle*{5}}
\put(265,97.5){\circle*{5}}
\end{picture}
\caption{Sketch of pedestrians moving through a cell of the obscure tunnel driven by a two-threshold biased dynamics.
\textit{Left panel}: the smaller black-filled circles represent the individuals located inside a cell, the bigger circles comprising the smaller black ones represent the interaction range of each individual. For an efficient communication network to be settled, a certain overlap among the bigger circles is needed, which is hence guaranteed by requiring the number of individuals in the cell to exceed the activation threshold $A$. \textit{Right panel}: as soon as the front row of the cell is full, the number of individuals occupying that front row, corresponding to the saturation threshold $S$, fixes an upper bound to the escape rate from the cell.}
\label{fig:threshold}
\end{figure}
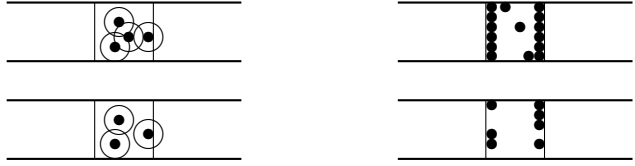

The rationale behind our modeling ideas fits the following 
{\em Gedanken experiment}. 
Imagine a flow of pedestrians  
on a lane and consider a partition of this lane in squared (or rectangular) cells; cf. Figure \ref{fig:threshold}.
The rate at which a walker leaves one cell is proportional 
to the number of pedestrian occupying the cell up to a limit
which is reached when the ``forward row'' of the cell is 
full, cf. the right panel of  Figure \ref{fig:threshold}. In this case, indeed, the pedestrians on the back are prevented from exiting the 
cell due to the presence of an obstacle. 
Thus, the escape rate from a cell increases proportionally to 
the number of pedestrians within the cell until this number reaches 
the total number of walkers that can be fit into the first row. 
On the other hand, the escape rate from a cell increases proportionally to the 
number of individuals provided that an efficient communication network (allowing the individuals to exchange 
informations about the location of the exit) can established inside the cell. Now, assuming that the interaction range, cf. the left panel of  
Figure \ref{fig:threshold}, between any pair of individuals is finite and much less than the size of the cell, the onset of an efficient communication network requires the 
number of individuals to exceed a minimal value which allows a proper interaction inside the cell. 

These effects are captured by using ZRP with, respectively, a saturation 
and an activation threshold \cite{CCM2016jnet}. In essence,  our modelling is 
rather simple: no interaction between pedestrians on different 
cells is taken into account. This choice is deliberate -- we want to 
keep the level of modelling as low as possible to show that, 
even in such cases, it is possible to recover the 
qualitative behavior of the fundamental diagrams. 

In the particular ZRP introduced in this paper, 
the two thresholds can be tuned so as to switch from an independent motion 
of the particles to a motion that can be mapped to a simple exclusion process. 
When considering the hydrodynamic limits of our model [(i) reversible dynamics, (ii) dynamics with a drift], the resulting macroscopic dynamics exhibit a non--trivial dependence on the thresholds, which is, to our knowledge, yet unexplored. 
 
The motivation for this study stems from our interest in the motion of pedestrian flows in dark or in heavily obscured corridors, where the internal dynamics of pedestrians  can change depending on the willingness to cooperate (here: to adhere to large groups) or to be selfish (here: to perform independent random walks); see \cite{CM01,CM02,CM03} for more details in this direction. 

To be able to understand the behavioral change leading individuals from cooperation to selfishness and eventually backwards, we thus opted for the introduction of two thresholds affecting the microscopic dynamics of the particle system. From the evacuation point of view, the central question is: 
{\em Which values of the thresholds yield higher evacuation fluxes (currents), or, in other words, allow for lower (average) residence times?} 
It is worth noting that this particular traffic scenario is intimately related to the dynamics of molecular motors seen from the perspective of  
processivity (cf., e.g., \cite{LH93}). For transporting at molecular scales,  one distinguishes between processive and non--processive motors. The processive ones  perform
best when working in small groups (porters), while  the non--processive motors work best in large groups (rowers). Their joint collective dynamics has been investigated in \cite{Joa06}. If the motors suddenly change their own processivity from porters to rowers (for instance, due to particular environmental conditions, or due to a command control from a hierarchical structure), then our approach based on zero range processes with threshold approximates conceptually well the changing--in--processivity dynamics. 

Threshold effects are not new in microscopic dynamics. They are usually introduced to model dynamics undergoing 
   sudden changes when some dynamical observable exceeds an {\em a priori}
   prescribed value. A natural application of this point of view appears in the context of infections propagation models, where an individual 
   gets infected if the number of infected neighbors is large enough. 
   A very well--studied situation is the Bootstrap Percolation 
   problem \cite{CLR} in which, for instance, on a square lattice, a site 
   becomes infected as soon as the number of its neighboring 
   infected sites is larger than a fixed threshold value. In this 
  context, the most interesting and surprising situation is the 
   one in which the threshold is precisely half of the total neighboring 
   sites. In such a case, new scaling laws have been discovered 
   in the infinite volume limit \cite{AL,CC}. 
   




In the next sections we will focus on the hydrodynamic limit of our ZRP built on thresholds, subjected to periodic boundary conditions and equipped with either symmetric or asymmetric jump probabilities. The asymmetry in the jump probabilities breaks the condition of detailed balance and gives hence rise to a net particle current across the system. We will explicitly highlight the effect of the thresholds (microscopic information) on the macroscopic transport equations
and discuss, in particular, the dependence of the structure of the effective diffusion coefficient and of the effective current on both the thresholds and the local pedestrian density. Our analysis allows one to recover some known results available for the independent particle model and for the simple exclusion process, and sets also the stage for a deeper understanding of the hydrodynamic limit of ZRP with a fixed number of thresholds. 

\section{The model}
\label{s:modello}
\par\noindent
We consider a positive integer $L$ and define a ZRP 
\cite{EH,pres} on the finite torus (periodic boundary conditions)
$\Lambda:=\{1,\dots,L\}\subset\bb{Z}$.
We fix $N\in\bb{Z}_+$ and consider the finite
\emph{state space}
$\Omega_{L,N}$:
\begin{equation}
\label{mod000}
\Omega_{L,N}
=
\Big\{
\omega\in\{0,\dots,N\}^\Lambda:\,
\sum_{x=1}^L\omega_x=N
\Big\}
\,\,
\end{equation}
where the integer $\omega_x$ denotes the \emph{number of particles}
at the site $x\in\Lambda$ in the \emph{state}
$\omega$.

We pick $A,S\in\{1,\dots,N\}$ with $S\ge A$, 
the \emph{activation} and \emph{saturation thresholds}, respectively.
We define, next, the \emph{intensity function}
\begin{equation}
\label{soglia}
g(k)
=
\left\{
\begin{array}{ll}
0 &  \textrm{ if } k=0\\
1 & \textrm{ if } 1\leq k\le A\\
k-A+1 & \textrm{ if } A< k\le S\\
S-A+1 & \textrm{ if } k> S\\
\end{array}
\right.
\end{equation}
for each $k\in\bb{Z}_+$. The intensity function, and all the quantities 
that we shall define below, do depend on the two thresholds $A$ and $S$, 
but we skip them from the notation for simplicity. 

The ZRP considered in this paper
is the Markov process $(\omega_t)_{t\ge0}\in\Omega_{L,N}$,
such that each 
site $x\in\Lambda$ is updated with intensity $g(\omega_x(t))$
and, once such a site $x$ is chosen, a particle jumps with 
probability $p\in[0,1]$ to the 
neighboring right site $x+1$ or with probability $1-p$ 
to the neighboring left site $x-1$ 
(recall periodic boundary 
conditions are imposed). For more details 
we refer the reader to \cite{pres,KL}.
In our model, the intensity function is related to the 
(time dependent, in general) \textit{hop rates} 
\begin{displaymath}
r^{(x,x-1)}(\omega_x(t))
=
(1-p)g(\omega_x(t))
\;\;\textrm{ and }\;\;
r^{(x,x+1)}(\omega_x(t))
=
pg(\omega_x(t))
\end{displaymath}
and coincides, hence, with the \textit{escape rate} 
$
r^{(x,x-1)}(\omega_x(t))
+
r^{(x,x+1)}(\omega_x(t))
=g(\omega_x(t))
$
at which a particle leaves the site $x$.

Thus, the effect of the thresholds is to control the 
escape rate from the site. More precisely, the activation threshold $A$ 
keeps the escape rate low and fixed to unity for all sites for 
which $\omega_x(t)\le A$, regardless of the number of particles on $x$.
The saturation threshold $S$, instead,  
holds the escape rate fixed to a maximum value for all sites for 
which $\omega_x(t)\ge S$, regardless, again, of the number of particles on $x$.
In the intermediate case, $A<\omega_x(t)<S$
the escape rate increases proportionally to the actual number of particles on 
$x$, see \eqref{soglia}. 

We remark that in the limiting case $A=1$ and $S=N$, 
the intensity function becomes  
$g(k)=k$, for $k>0$, hence
the well known
\emph{independent particle} model is recovered.
A different limiting situation is the one in which 
the intensity function is set equal to $1$ for any $k\ge1$ and 
equal to zero for $k=0$.
In this case, the configurations of the ZRP can be mapped to the 
simple exclusion model states, see e.g. \cite{EH}, and we shall thus refer to the latter case as the 
\emph{simple exclusion}--like model. Such a model is 
found, in our set--up, when $A=S$.
We point out that one of the interesting features of our model is the 
fact that it is able to tune between two very different dynamics, namely, 
the independent particle and simple exclusion--like behaviors \cite{EH}:
this tuning can be realized in two ways, i.e., 
by keeping $S=N$ and varying $A$ or by keeping $A=1$ and 
varying $S$. 

We are interested in studying
the hydrodynamic limit of this model, i.e. as $N\to \infty$ and $L\to \infty$. In particular we shall exploit 
the fact that the intensity function is not decreasing to 
use well established theories and derive in our set--up 
the limiting (effective) diffusion coefficient as well as the limiting  (effective) current in presence of the 
two thresholds. As we shall discuss later, the behavior of such 
macroscopic quantities with the local density will exhibit very 
peculiar features inherited from the microscopic properties of the 
dynamics. In particular, it will be possible to give a nice 
interpretation of the diagrams in terms of pedestrian motion, and 
the related fundamental diagrams will be explained in the framework 
of our very simple model. 

We let the \emph{Gibbs measure} with \emph{fugacity} $z\in\mathbb{R}$ 
of the ZRP introduced above 
be the product measure on $\mathbb{N}^\Lambda$ 
\begin{equation} 
\label{gibbs}
\prod_{x=1}^L \nu_{z}(\eta_x) 
\;\;\textrm{ for any }\;\;\eta=(\eta_1,\dots,\eta_L)\in\mathbb{N}^\Lambda
\end{equation}
with
\be 
\label{nu}
\nu_{z}(0)=C_{z} 
\;\;\textrm{ and }\;\;
\nu_{z}(k)=C_{z} \frac{z^k}{g(1)\cdots g(k)} 
\;\;\textrm{ for }k\ge1
\;\;, 
\ee
where $C_{z}$ is a normalizing factor depending, in general, 
on $z$, $A$, and $S$, namely, 
\begin{equation}
\label{norm000}
C_z
=
\Big[
     1+\sum_{k=1}^\infty\frac{z^k}{g(1)\cdots g(k)}
\Big]^{-1}
\;\;
\end{equation}


It is of interest to compute the mean value (against the 
Gibbs measure) of the intensity function $g$. By using \eqref{nu}, we get 
\begin{equation}
\label{mediaI}
\nu_{z}[g(\omega_x)]
=
\sum_{k=0}^\infty \nu_{z}(k)
\,g(k)
=
\sum_{k=1}^\infty \nu_{z}(k)
\,g(k)
=
C_{z} z + C_{z} z \sum_{k=2}^\infty \frac{z^{k-1}}{g(1)\cdots g(k-1)}
=z,
\end{equation}
where we have used that $g(0)=0$ and, in the last step, 
we recalled \eqref{norm000}.
 
We find relevant to stress that the expression of such an expectation, as 
a function of the activity, does not depend on the particular 
choice of the intensity function. 
Note, also, that the intensity is a site--dependent function,
whereas its expected value, with respect to the Gibbs measure given 
in \eqref{gibbs}, is not. This is due to the fact that 
the Gibbs measure is not site--dependent, which, in turn, stems from the 
imposed periodic boundary conditions and from the translationally invariant jump 
probabilities. 

\begin{figure}[h]
\centering
\includegraphics[width=7.7cm]{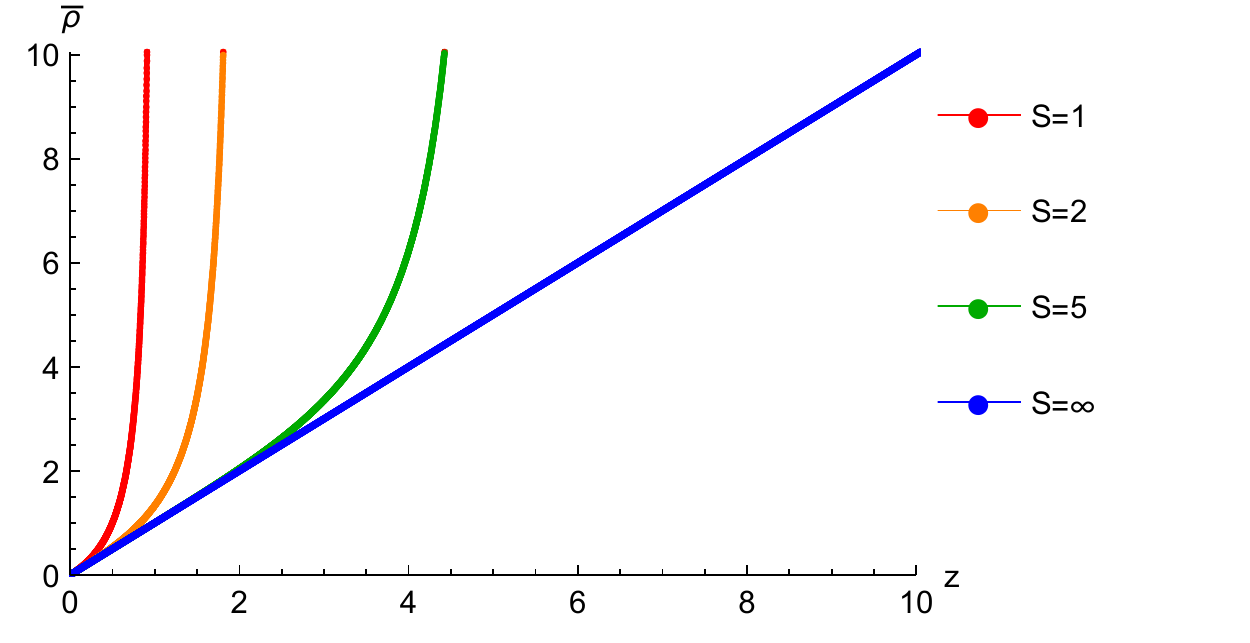}
\hspace{1mm}
\includegraphics[width=7.7cm]{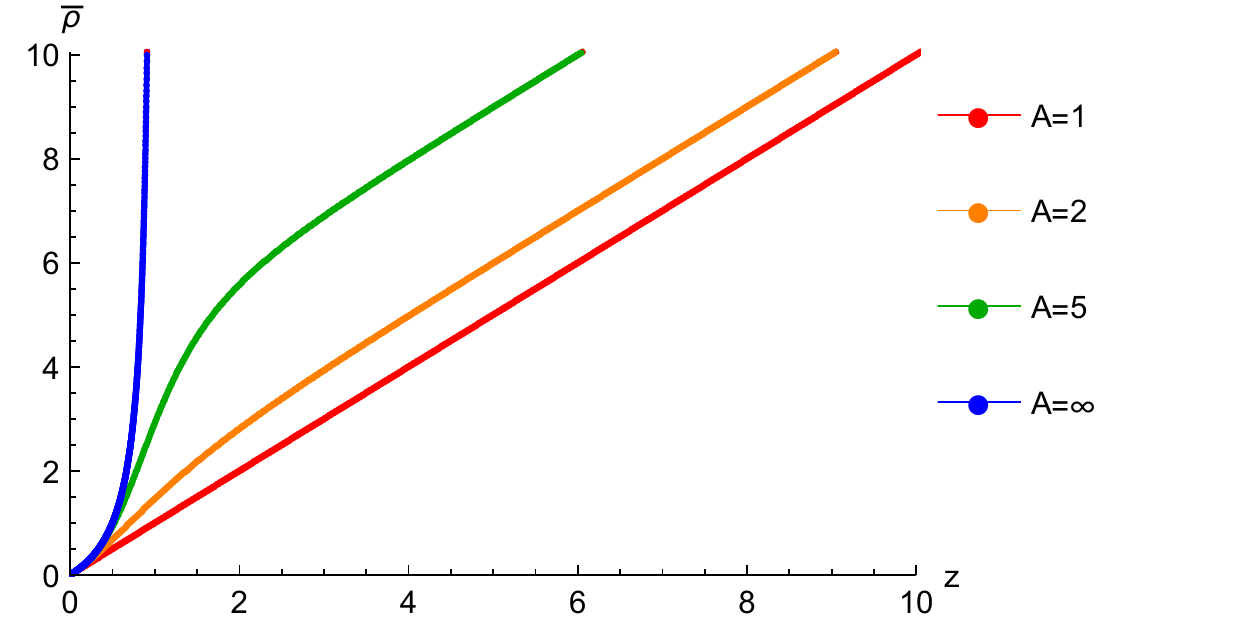}
\vspace{1mm}
\includegraphics[width=7.7cm]{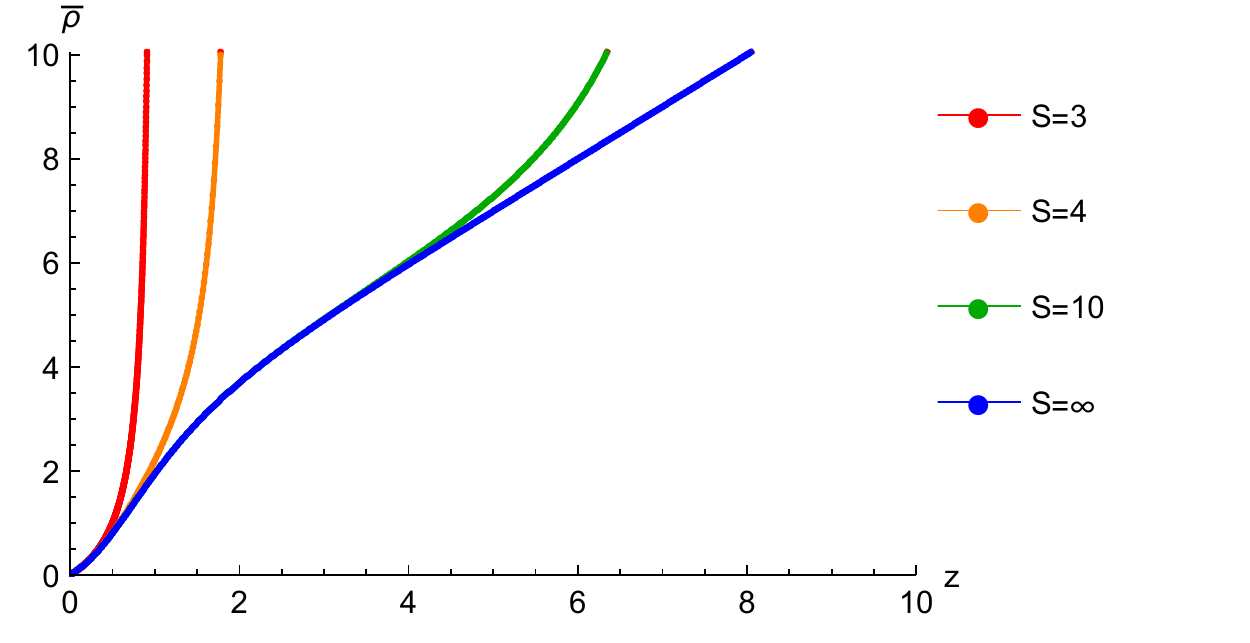}
\hspace{1mm}
\includegraphics[width=7.7cm]{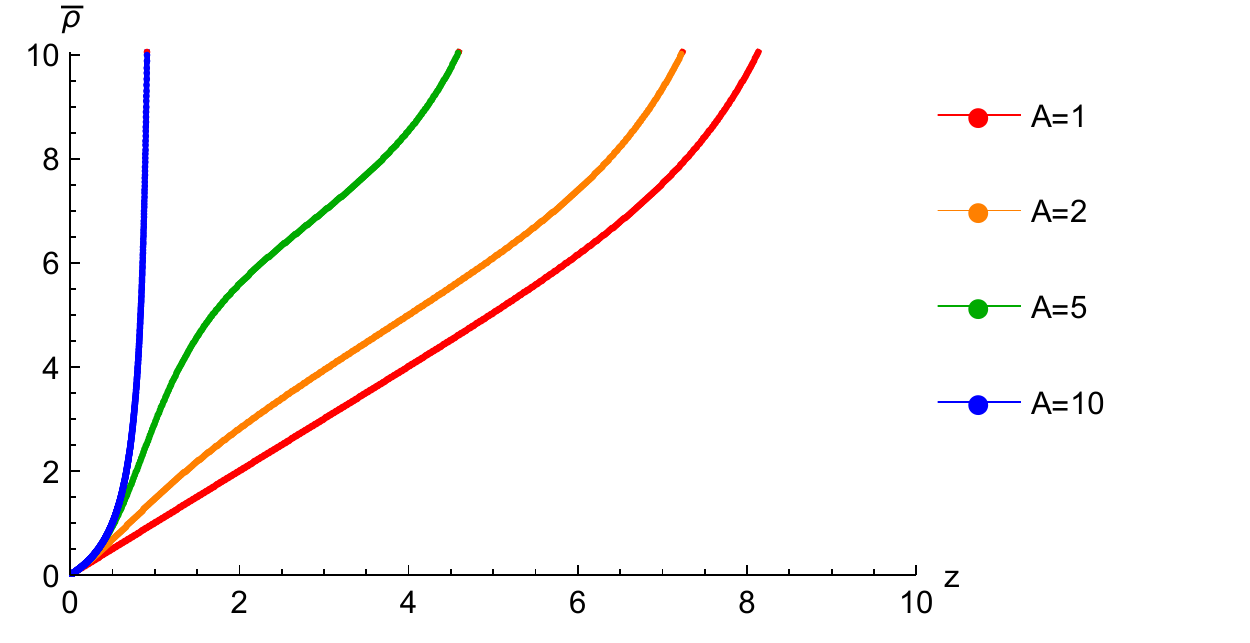}
\caption{\textit{Left panel, top row}: 
Graph of the function $\bar\rho(z)$ for 
$A=1$ and for different values of the saturation
threshold, i.e. $S=1,2,5,\infty$.
\textit{Right panel, top row}: 
Graph of the function $\bar\rho(z)$ for 
$S=\infty$ and for different values of the activation threshold, i.e. $A=1,2,5,\infty$. \textit{Left panel, bottom row}: Graph of the 
function $\bar\rho(z)$ for 
$A=3$ and for $S=3,4,10,\infty$. \textit{Right panel, bottom row}: 
Graph of the function $\bar\rho(z)$ for $S=10$ and for $A=1,2,5,10$.
}
\label{f:rhobar}
\end{figure}

In discussing the hydrodynamic limit, a special role is
played by the function
\be 
\label{dens}
\bar{\rho}(z)
=
\sum_{k=0}^\infty k\,\nu_{z}(k) 
\ee
It is possible to prove a nice expression for the function $\bar{\rho}$ 
independently of the particular choice of the intensity function. 
Indeed, recalling \eqref{nu}, equation \eqref{dens} can be rewritten as
\begin{displaymath}
\bar{\rho}(z)
=
C_{z}\sum_{k=1}^\infty k\,\frac{z^k}{g(1)\cdots g(k)} 
=
z\,C_{z}\frac{\textrm{d}}{\textrm{d}z}
   \sum_{k=1}^\infty \frac{z^k}{g(1)\cdots g(k)} 
=
z\,C_{z}\frac{\textrm{d}}{\textrm{d}z}
   \frac{1}{C_{z}}\,
   \sum_{k=1}^\infty \nu_{z}(k)
\end{displaymath}
which implies 
\begin{equation}
\label{dens02}
\bar{\rho}(z)
=
z\,C_{z}\frac{\textrm{d}}{\textrm{d}z}
   \frac{1}{C_{z}}
=
-\frac{z}{C_{z}}\,\frac{\textrm{d}}{\textrm{d}z}C_{z}
=
-z\,\frac{\textrm{d}}{\textrm{d}z}\log C_{z}
\;\;.
\end{equation}
At the same level of generality, 
it is not difficult to prove that
$\bar\rho(z)$ is an increasing function of the 
fugacity. Indeed, after some straightforward 
algebra, one can prove that 
\begin{equation}
\label{derivata}
\frac{\partial}{\partial z}\bar\rho(z)
=
\frac{\partial}{\partial z}
C_{z}\sum_{k=1}^\infty \frac{kz^k}{g(1)\cdots g(k)}
=
\frac{1}{z}[\nu_{z}(\eta_1^2)-(\nu_{z}(\eta_1))^2]
>0
\;\; .
\end{equation}
We mention that the above result is strictly connected to the fact 
that $(-\log C_{z})$ is a convex function. 

Finally, we observe that 
$\bar{\rho}$ is defined for any positive $z$ 
if $A$ is finite and $S=\infty$. On the other hand, it displays 
a singularity, i.e. it is defined for $z$ small enough, 
if $S$ is finite or when $A=S$ (simple exclusion--like model); 
see Figure~\ref{f:rhobar}.

\begin{figure}
\centering
\includegraphics[width=7.7cm]{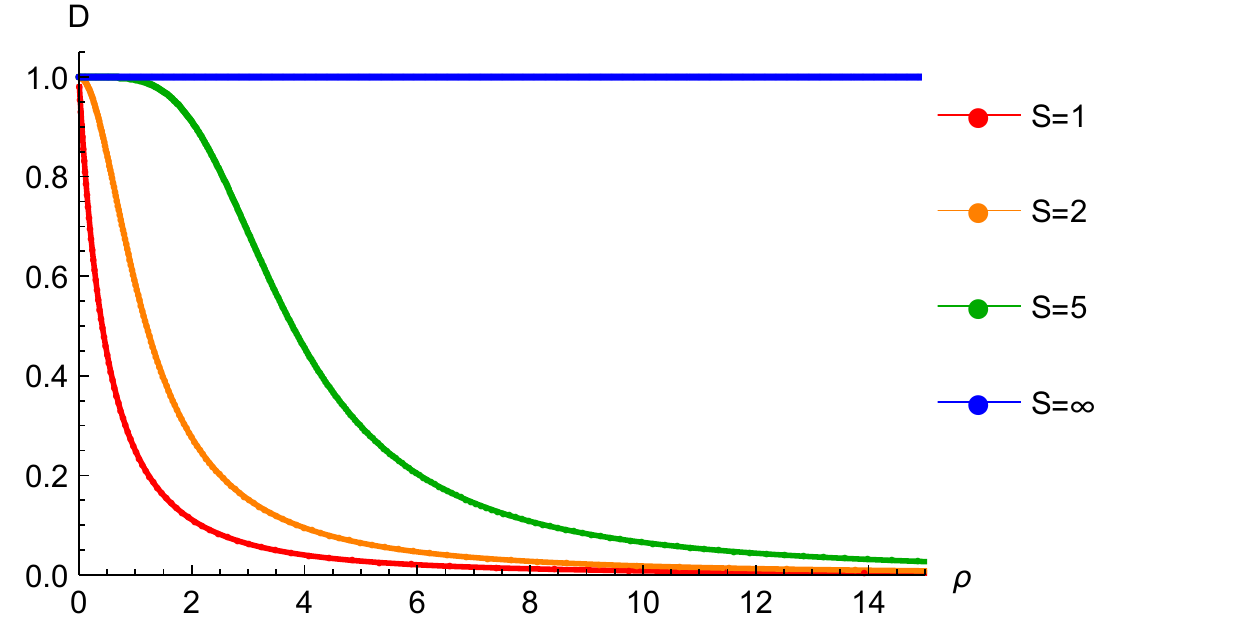}
\hspace{1mm}
\includegraphics[width=7.7cm]{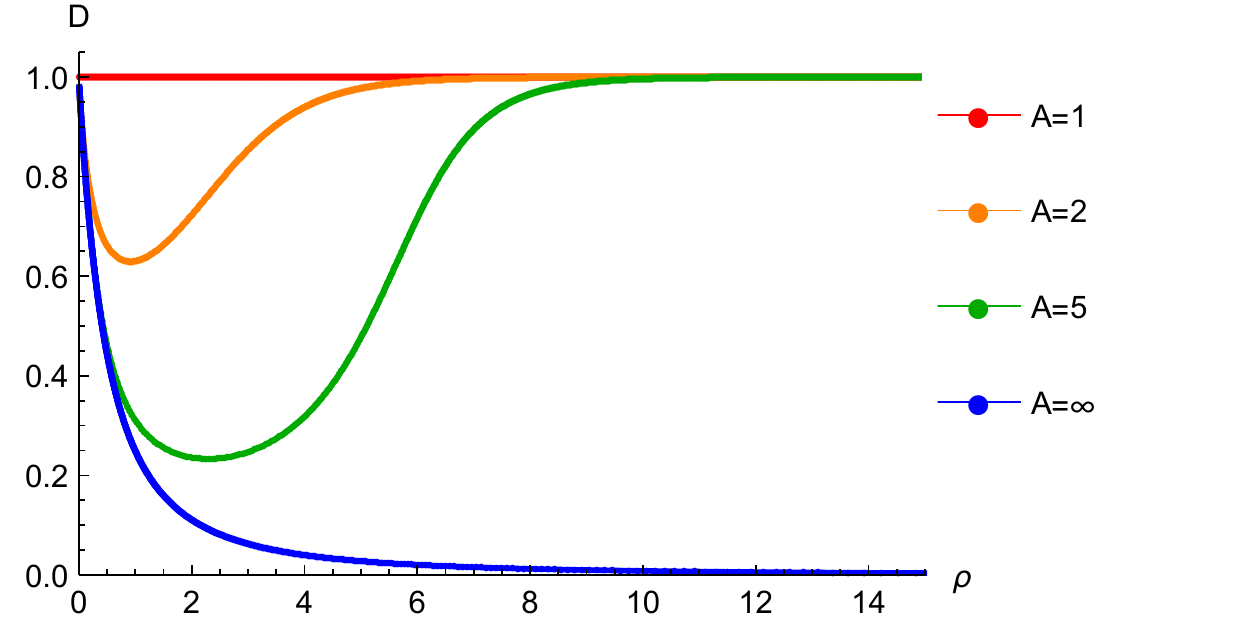}
\vspace{1mm}
\includegraphics[width=7.7cm]{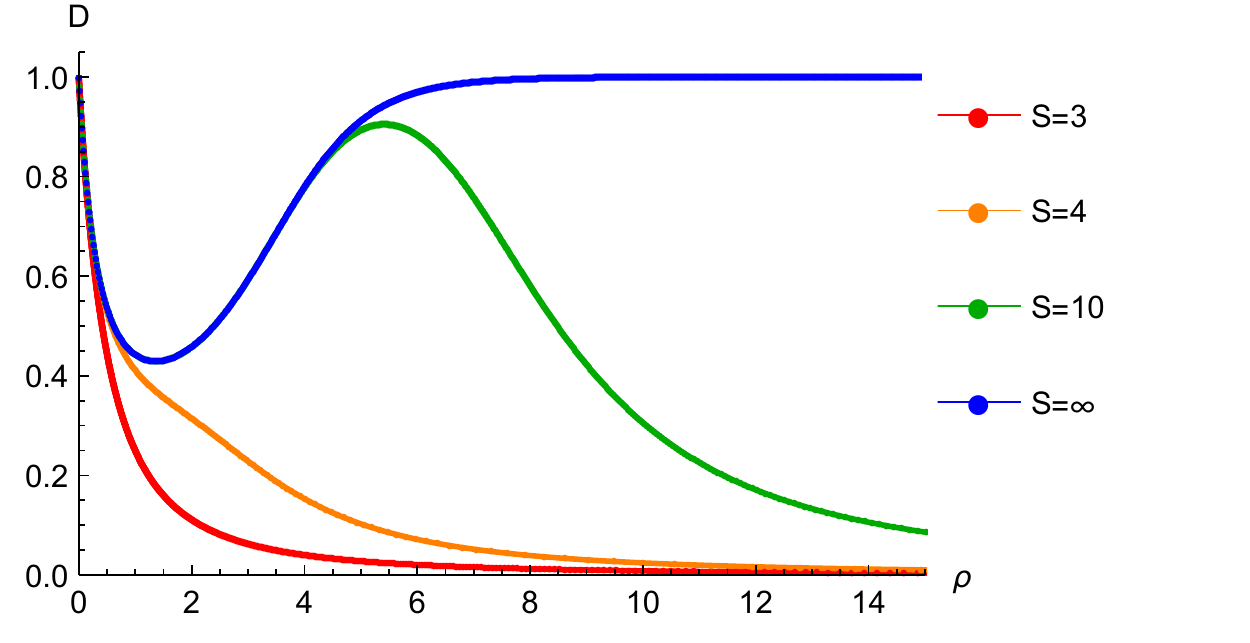}
\hspace{1mm}
\includegraphics[width=7.7cm]{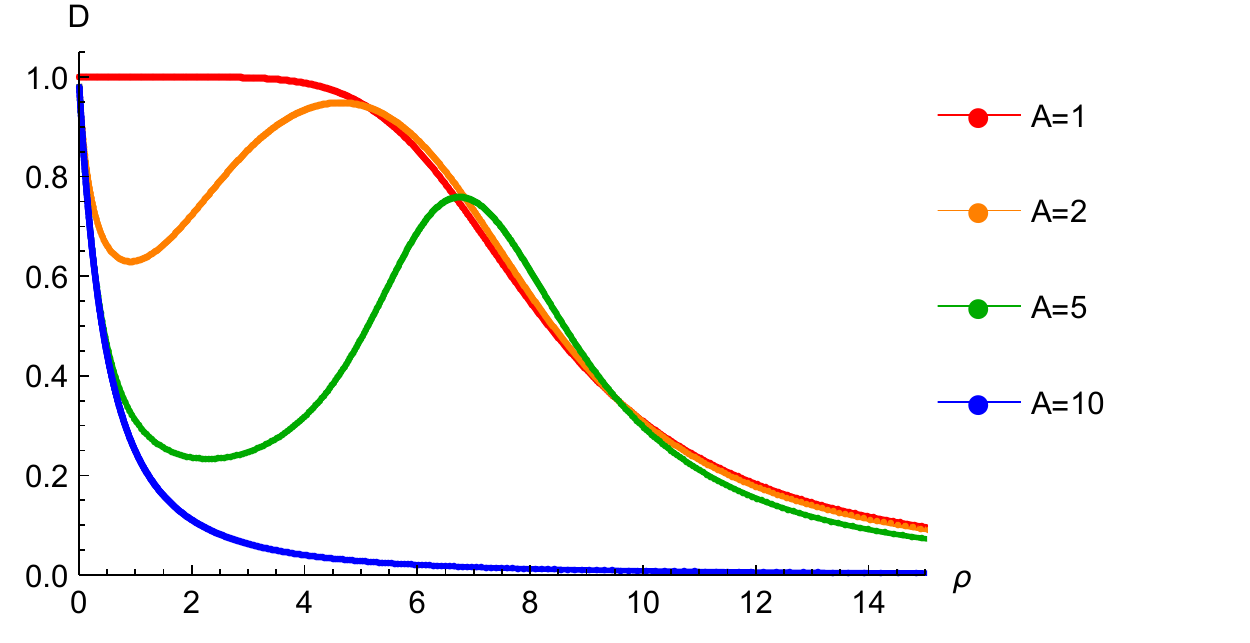}
\caption{\textit{Left panel, top row}: Behavior of the diffusion coefficient 
$D(\rho)$ vs. $\rho$ for $A=1$ and for different values of the 
saturation threshold, 
i.e., $S=1,2,5,\infty$. 
\textit{Right panel, top row}: Behavior of the diffusion coefficient 
$D(\rho)$ vs. $\rho$ for 
$S=\infty$ and for different values of the activation threshold, 
i.e., $A=1,2,5,\infty$. 
\textit{Left panel, bottom row}: Behavior of the diffusion coefficient 
$D(\rho)$ vs. $\rho$ for $A=3$ and for 
$S=3,4,10,\infty$. \textit{Right panel, bottom row}: Behavior of the diffusion coefficient 
$D(\rho)$ vs. $\rho$ for $S=10$ and for $A=1,2,5,10$.}
\label{fig:diff}
\end{figure}

\section{Hydrodynamic limit for reversible dynamics}
\label{s:hydlimeq}
\par\noindent
The dynamics with $p=1/2$ is
{\em reversible} with respect to the invariant measure. 
The evolution of the distribution of the particles 
on the space $\Lambda$ for the ZRP
with thresholds $A$ and $S$ introduced above can be described in the diffusive 
hydrodynamic limit via the time evolution of the \textit{density 
function} $\rho(x,t)$, with the space variable $x$ 
varying in the interval $[0,1]$ and $t\ge0$. 

In the framework of one--dimensional ZRP, 
hydrodynamic equations are derived rigorously under 
the assumption that the intensity function is not decreasing.
We refer to \cite[Chapter~III]{pres} and 
\cite[Chapter~5]{KL} 
for a detailed discussion and the rigorous proof. 
The first proof of this result can be found in \cite{DMF}
and is based on the results reported in \cite{Andjel}. 

It suffices, here, to recall the main findings: 
one can prove that for $p=1/2$ the continuous 
space density $\rho(x,t)$ 
is the solution of the partial differential equation 
\be 
\label{diffusion}
\frac{\partial}{\partial t}\rho
=
-\frac{\partial}{\partial x}{J(\varrho)} 
\ee
where the \textit{macroscopic flux} $J(\varrho)$ is defined as
\be 
\label{J}
J(\varrho)
=
-\frac{1}{2}
D(\rho)
\frac{\partial}{\partial x}\rho
\ee
with the \textit{diffusion coefficient} $D$ given by
\be 
\label{D}
D(\rho)
=
\frac{\partial }{\partial \rho}
\nu_{\bar{z}(\rho)}\left[g(\omega_1)\right] 
\;\;.
\ee
Note that the diffusion coefficient is here computed in terms 
of the mean of the intensity function evaluated against the 
single site Gibbs measure with fugacity corresponding to the 
local value of the density. 

Note that, even if it is not coded in the notation, 
the diffusion coefficient $D$  depends on the values of the thresholds.
One of the main multi-scale aspects of our analysis is, indeed, precisely the link between the two thresholds $A$ and $S$ and the effective diffusion coefficient $D$.

We shall first recall the well known results which hold in the 
limiting cases corresponding to the independent particles and simple exclusion--like 
dynamics. 

\br
\label{rem1}
\textit{Independent particle model:\/}
For $A=1$ and $S=\infty$, one has 
$C_z=\exp\{-z\}$. Hence, by \eqref{dens02}, it holds
$\bar{\rho}(z) = z$.
Thus, 
recalling \eqref{mediaI} and 
the definition of $\bar{z}$ given 
below \eqref{derivata}, 
one finds 
$\nu_{\bar{z}(\rho)}\left[g(\omega_1)\right] 
=
\nu_{\rho}\left[g(\omega_1)\right] 
=
\rho
$.
Thus, by using \eqref{D},
the diffusion coefficient reads
$D(\rho)
=1$.
\er

\br
\label{rem2}
\textit{Simple exclusion--like model:\/}
For $A=S$ (either finite or infinite), 
one has $g(k)=1$ for any $k\ge1$ and $g(0)=0$.
Hence, $C_z=1-z$, and it holds 
$\bar{\rho}(z) = z/(1-z)$.
Thus, proceeding as above, one finds the law
$D(\rho)=1/(1+\rho)^2$, cf. \cite{Ferrari}.
\er

Hence, in the two limiting cases, one can easily determine the expression of the diffusion coefficient. In the general case, i.e.
for arbitrary values of the thresholds $A$ and $S$, we 
exploit the following strategy. 
We use, first, \eqref{norm000} and \eqref{dens02} to compute 
$\bar{\rho}(z)$, whose explicit expression in terms of special 
functions is reported in Appendix~\ref{sec:app1}.
Then, we compute the diffusion coefficient via the equation \eqref{D}, 
where we use equation \eqref{mediaI} to express the average 
of the intensity function with respect to the Gibbs measure and invert 
the function $\bar{\rho}(z)$ to obtain $\bar{z}(\rho)$. 
More concisely, we write 
\begin{equation}
\label{Dgen}
D(\rho)
=
\frac{\partial}{\partial\rho}
\nu_{\bar{z}(\rho)}\left[g(\omega_1)\right] 
=
\frac{\partial}{\partial\rho}
\bar{z}(\rho)
=
\Big(
\frac{\partial}{\partial z}
\bar{\rho}(z)
\Big)^{-1}
\Big|_{z=\bar{z}(\rho)}
\end{equation}
We remark that the explicit expression of the 
quantity 
$\partial\bar\rho(z)/\partial z$ appearing in 
\eqref{Dgen} is quite 
lengthy and will be omitted here.
By performing the above computation, we thus obtain the expression 
of the diffusion coefficient $D(\rho)$.

Figure~\ref{fig:diff} shows the behavior of the diffusion 
coefficient as a function of the local density
and parameterized by the values 
of the thresholds. In particular, the upper left panel of 
Figure~\ref{fig:diff} refers to the case $A=1$ 
and for different values of $S$: the simple exclusion--like model is recovered for $S=1$ , while the independent particle model is attained for 
$S=\infty$. Similarly, the upper right panel illustrates the case with 
$S=\infty$ 
and for different values of $A$: here the independent particle model corresponds to $A=1$ and the simple
exclusion--like model is found for $A=\infty$. As shown in both the upper panels of Figure~\ref{fig:diff}, in the independent particle 
case the diffusion coefficient is 
constant with respect to the local density and is equal to unity. 

A noteworthy feature of the diffusion coefficient, clearly visible in the upper right panel as well as in both the lower panels of Figure~\ref{fig:diff},
is the loss of monotonicity of the function $D(\rho)$ occurring at values of $\rho$ exceeding 
some critical value (depending, in general, on $A$ and $S$). 
This remark can be interpreted 
as the effect, 
at the hydrodynamic level, of an activation threshold $A>1$ and/or 
$S< \infty$ acting at the more microscopic, dynamical, level: both conditions locally pull 
the dynamics
away from the independent particle behavior.

Note, for instance, the behavior of $D(\rho)$ displayed in the lower left panel of Figure~\ref{fig:diff}, 
referring to the case $A=3$.
Considering, in particular, the green curve, corresponding to 
$S=10$, one observes 
the onset of a double loss of monotonicity of the function $D(\rho)$: 
for small values of the density, $D$ stays close to the simple exclusion--like behavior and 
decreases with $\rho$, 
then, after one first critical value of the density, it starts rising up, until it eventually 
drops down again, when $\rho$ exceeds an upper critical value.
This reflects precisely the existence of a double 
threshold for the intensity function, 
described by \eqref{soglia}.
More precisely, if the local density is smaller than some 
critical value (close to the activation threshold $A$) the behavior is 
essentially simple exclusion--like, because the intensity function 
is fixed to unity for the typical values of the number of on site particles 
corresponding to such a density. On the other hand, 
if the local density exceeds this first critical value, the typical 
number of on site particles happens to fall above the activation 
threshold. Hence, since in this regime the intensity function 
is proportional to the number of on site particles, the diffusion 
coefficient starts growing as a function of the local density. 
Finally, if the local density exceeds a second critical value 
(close to the saturation threshold $S$), the intensity function 
attains a constant value independently on the number of on site 
particles, and the diffusion coefficient behaves, as it again pertains to the simple exclusion--like regime, as a decreasing 
function of the local density. 

The effect of the two thresholds on the diffusion coefficient is, therefore, clear: 
at fixed saturation threshold, the diffusion coefficient decreases with
increasing activation threshold. 
On the other hand, at fixed activation threshold , 
the diffusion coefficient increases with
increasing saturation threshold. 
Moreover, in presence of reversible dynamics,
the dependence of the diffusion coefficient with respect to 
density may become non--monotonic.



\begin{figure}
   \centering
   \includegraphics[width=6cm]{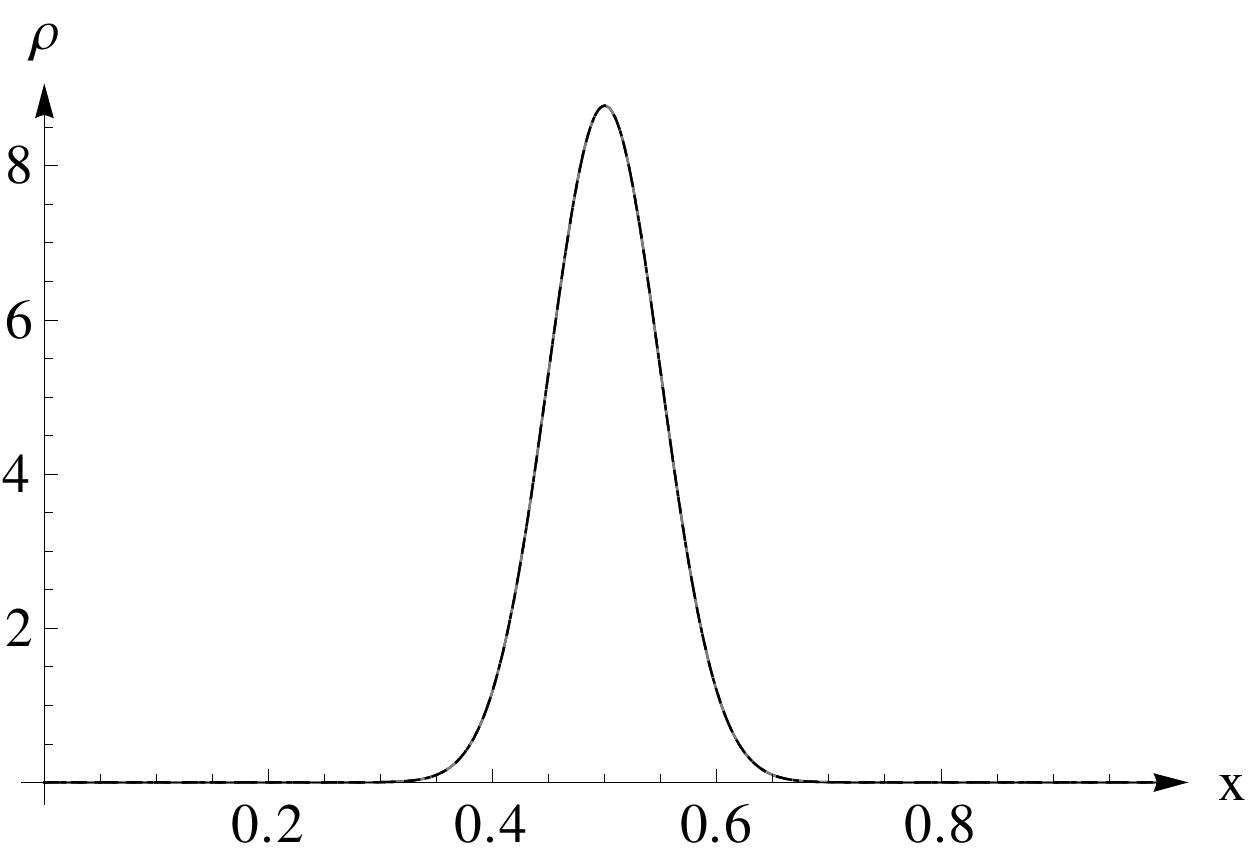}
   \includegraphics[width=6cm]{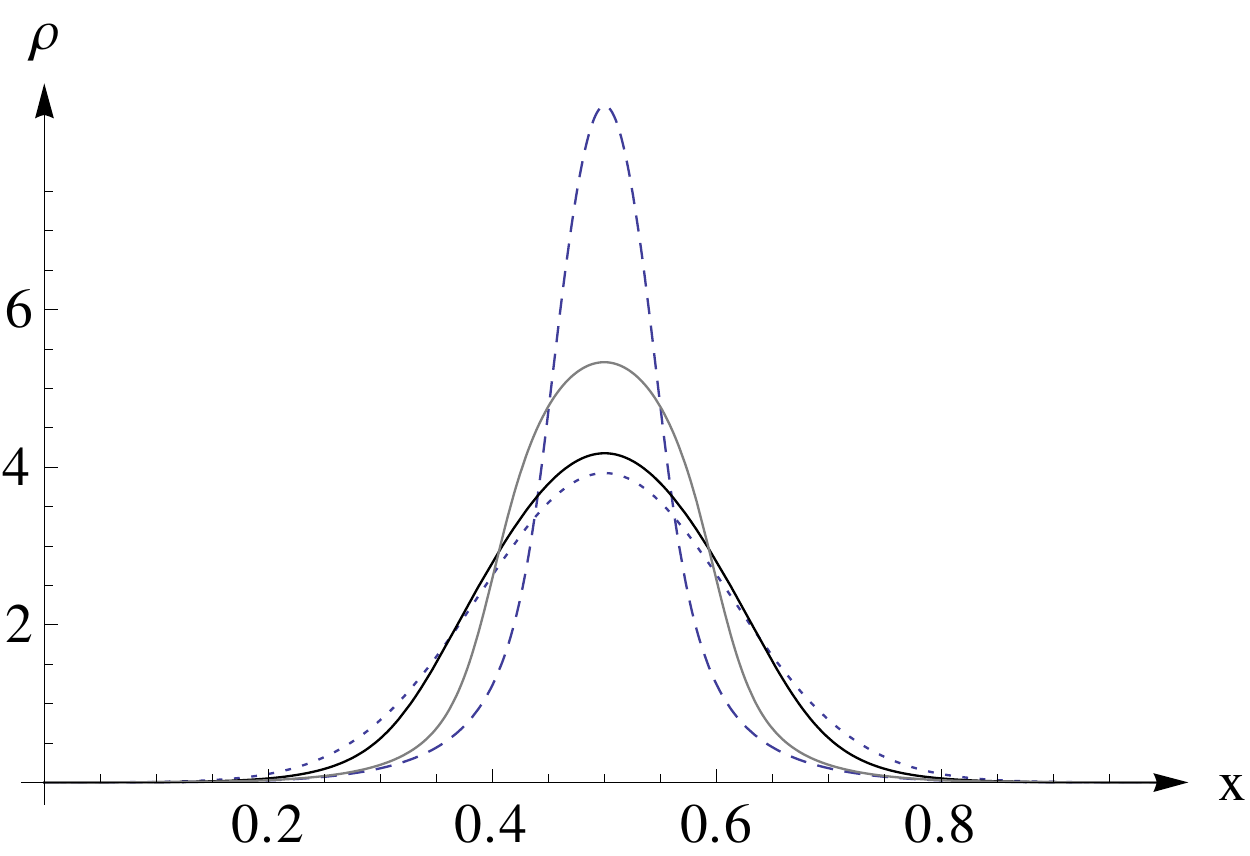}
\\
$\phantom.$
\\
   \includegraphics[width=6cm]{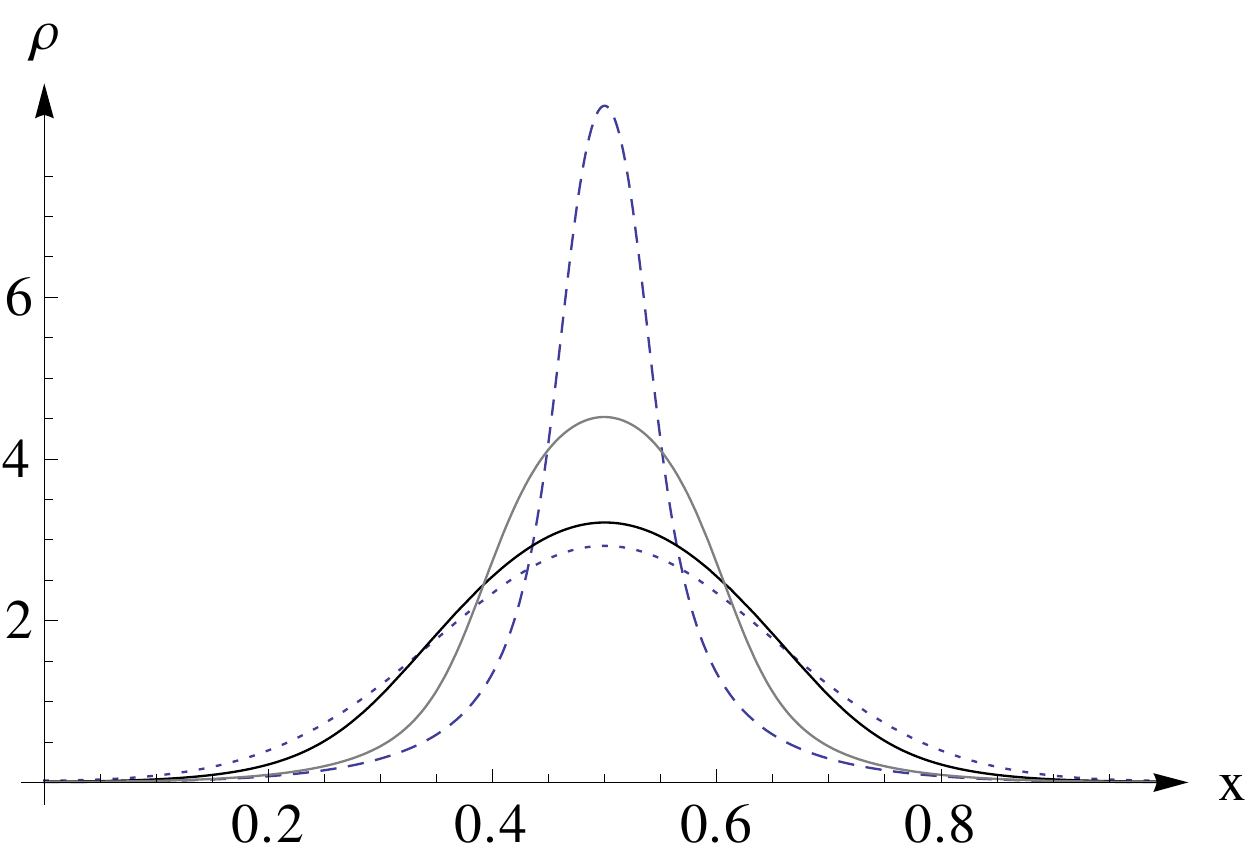}
   \includegraphics[width=6cm]{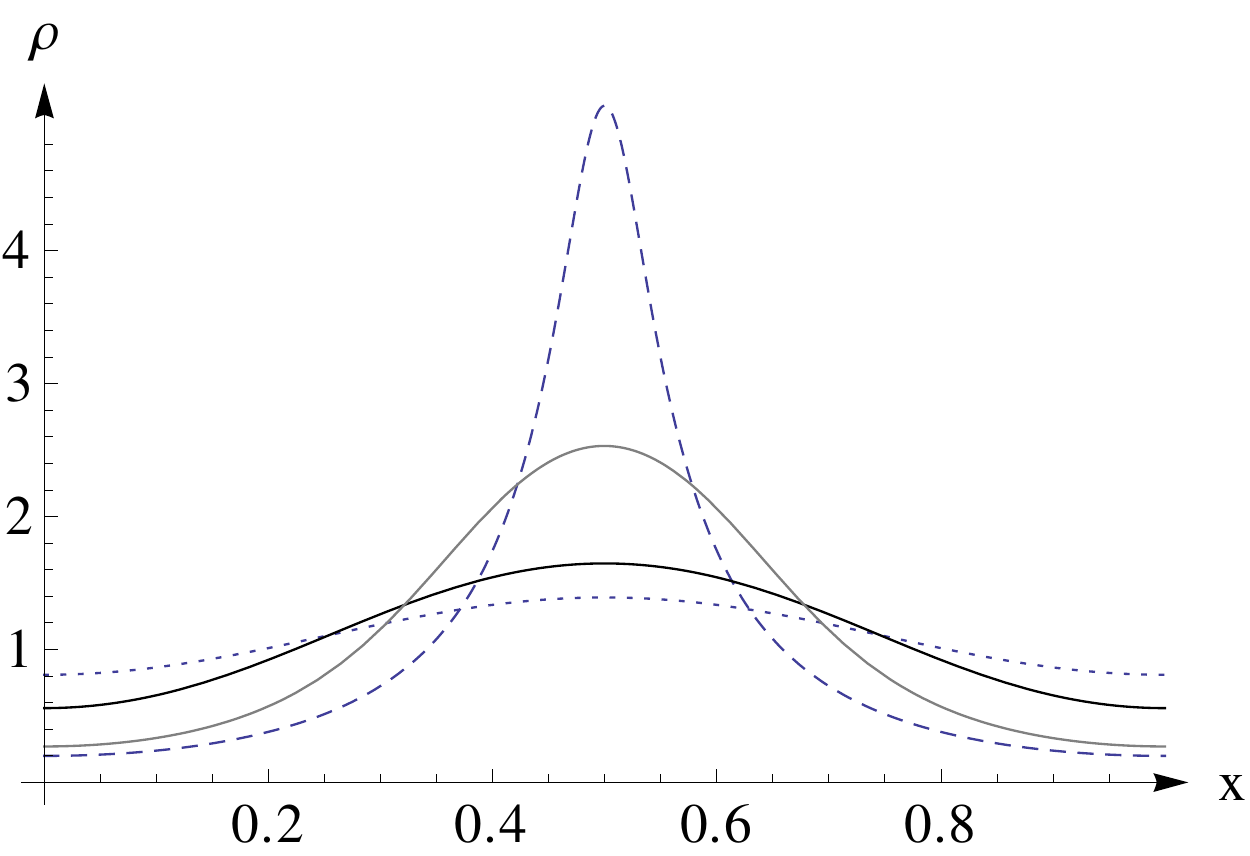}
\\
   \caption{Plot of the solution of the equation \eqref{diffusion} $\rho(x,t)$ 
vs.\ $x$, with a given initial condition and with periodic boundary conditions.
Dotted and dashed lines refer to the two limiting cases corresponding, 
respectively, to the independent particle (Remark~\ref{rem1}) 
and to the simple exclusion--like (Remark~\ref{rem2}) process.
Solid lines refer to intermediate cases 
$A=5$ and $S=10$ (gray) and $A=2$ and $S=10$ (black). 
Different panels, in lexicographic order, 
report data referring to time 
$t=0$, $t=0.01$, $t=0.02$, $t=0.1$.}
\label{pde}
\end{figure}

\begin{figure}
   \centering
   \includegraphics[width=6cm]{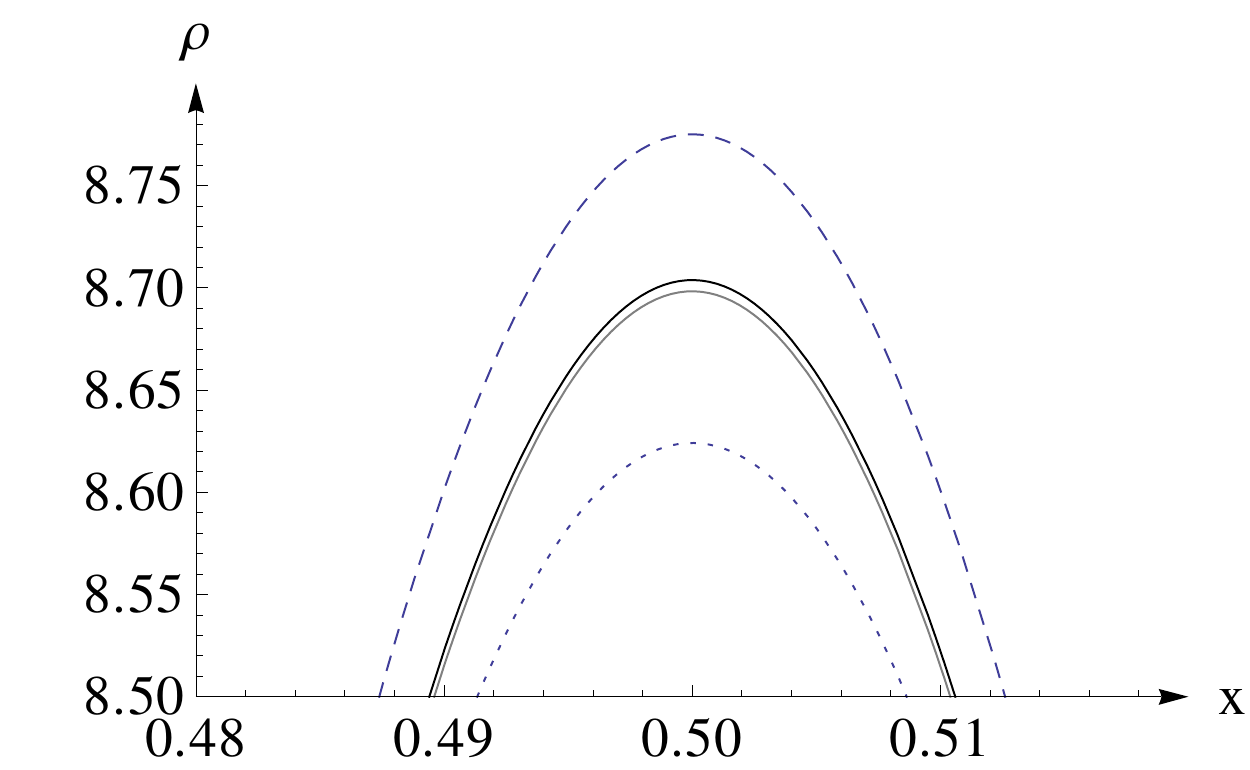}
   \includegraphics[width=6cm]{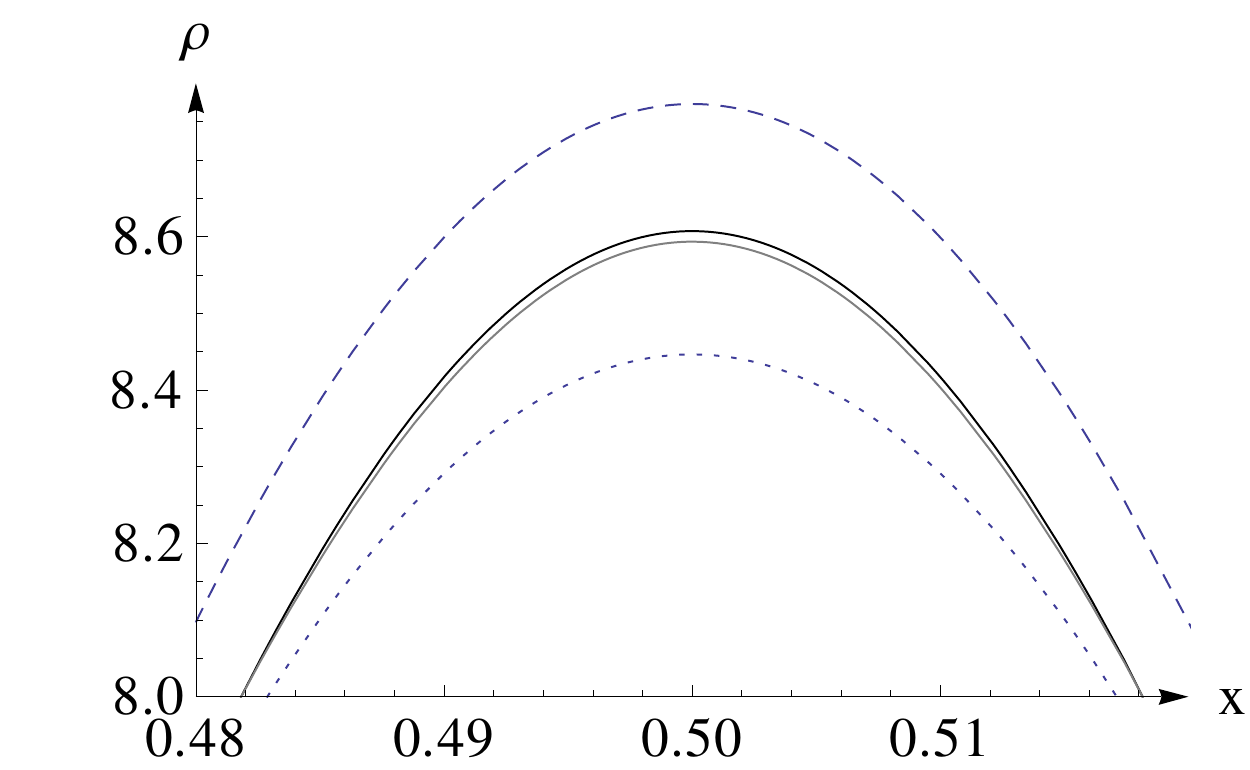}
\\
$\phantom.$
\\
   \includegraphics[width=6cm]{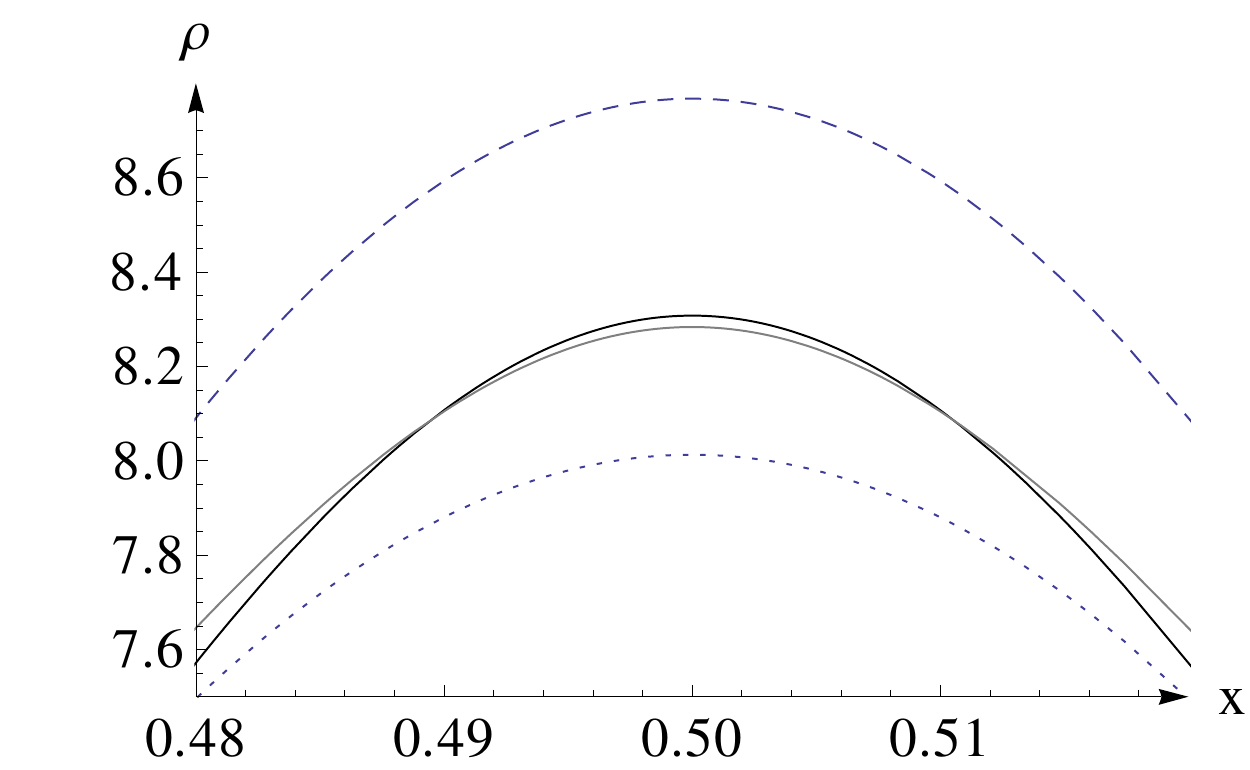}
   \includegraphics[width=6cm]{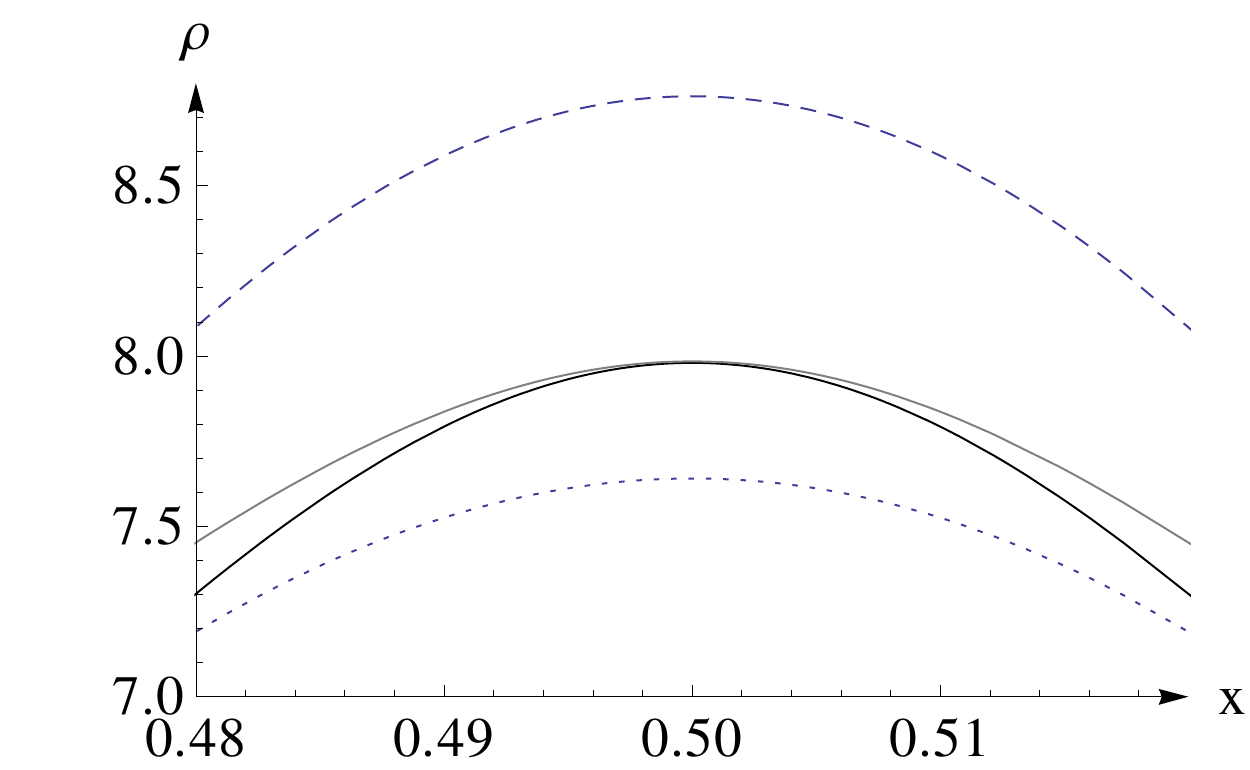}
\\
   \caption{The same as in Figure~\ref{pde}.
Different panels, in lexicographic order, 
report data referring to time 
$t=0.0001$, $t=0.0002$, 
$t=0.0005$, $t=0.0008$.}
\label{pde02}
\end{figure}

The resulting behavior of the diffusion coefficient can be better understood by also 
recalling that the number $\mathcal{N}(t)$ of particles departing from the site $x\in[0,1]$, is described by a 
non--homogeneous Poisson process with time--dependent rate parameter 
$g(\omega_x(t))$ (the escape rate). Thus, given a small $\delta>0$, it holds
$$
P_t[\mathcal{N}(\delta)=1]\simeq g(\omega_x(t)) \delta
$$
where $P_t[\mathcal{N}(\delta)=1]$ is the probability of 
exactly one change in $\omega_x(t)$ in the time interval 
$(t,t+\delta)$. Then, for values of the threshold $A$ and 
$S$ different, respectively, from $1$ and $\infty$ (independent particle model),   
$g(\omega_x)$ takes a lower value compared to that 
referring to the independent particle model, 
with a minimum (corresponding to $g(\omega_x)=1$) attained when $A=S$ 
(i.e., simple exclusion--like model).

The effect of the threshold on the dynamics, in the hydrodynamic limit, 
is also visible in Figure~\ref{pde}, 
showing the profiles, at different times, of the function 
$\rho(x,t)$ solving \eqref{diffusion}--\eqref{J}, for four different choices of the thresholds.
The numerical solutions of the PDE \eqref{diffusion} 
exhibit the fastest decay in the independent particle case and 
the slowest one in the simple exclusion--like case. 
Whereas in the two other plotted cases the decay rate 
is intermediate. 
This is in perfect agreement with the data plotted for the diffusion 
coefficient in Figure~\ref{fig:diff}: indeed, 
such a coefficient is maximal in the independent particle case and 
minimal in the simple exclusion--like situation. 

Similar data, at different times, have been plotted in 
Figure~\ref{pde02}.
We note that the curves corresponding to the two cases 
$A=5$ and $S=10$ (gray) and $A=2$ and $S=10$ (black)
swaps while time goes by.
This behavior is in perfect agreement with that 
shown by the diffusion coefficient in 
the bottom right panel of Figure~\ref{pde02}.

\section{Hydrodynamic limit in presence of a drift}
\label{hydlimdrift}
In Section \ref{s:hydlimeq} we discussed the effect of the thresholds 
on the diffusion equation describing the macroscopic behavior of the 
system in the hydrodynamic limit. In this Section we investigate 
how the dynamics depends on the thresholds under the effect of an 
external field breaking the condition of detailed balance and
inducing a non--vanishing particle current across the system. 
That is, we tackle, here, the analysis of the hydrodynamic limit of the ZRP with 
$p\neq1/2$ and in presence of the two thresholds. 

The evolution of the distribution of the particles 
for a ZRP 
subjected to the two aforementioned thresholds 
and to a non--vanishing drift can be described, in the hydrodynamic limit, in terms of the \textit{density 
function} $\rho(x,t)$ with the space variable $x$ 
varying in the interval $[0,1]$ and $t\ge0$. 

\begin{figure}
\centering
\includegraphics[width=7.7cm]{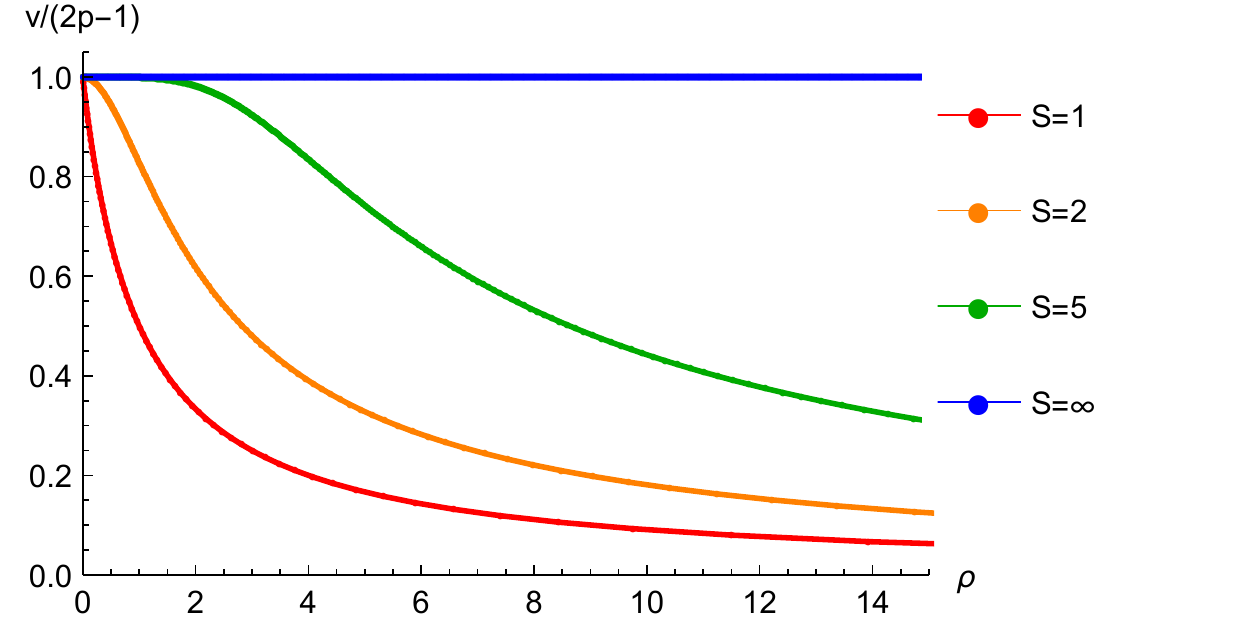}
\hspace{1mm}
\includegraphics[width=7.7cm]{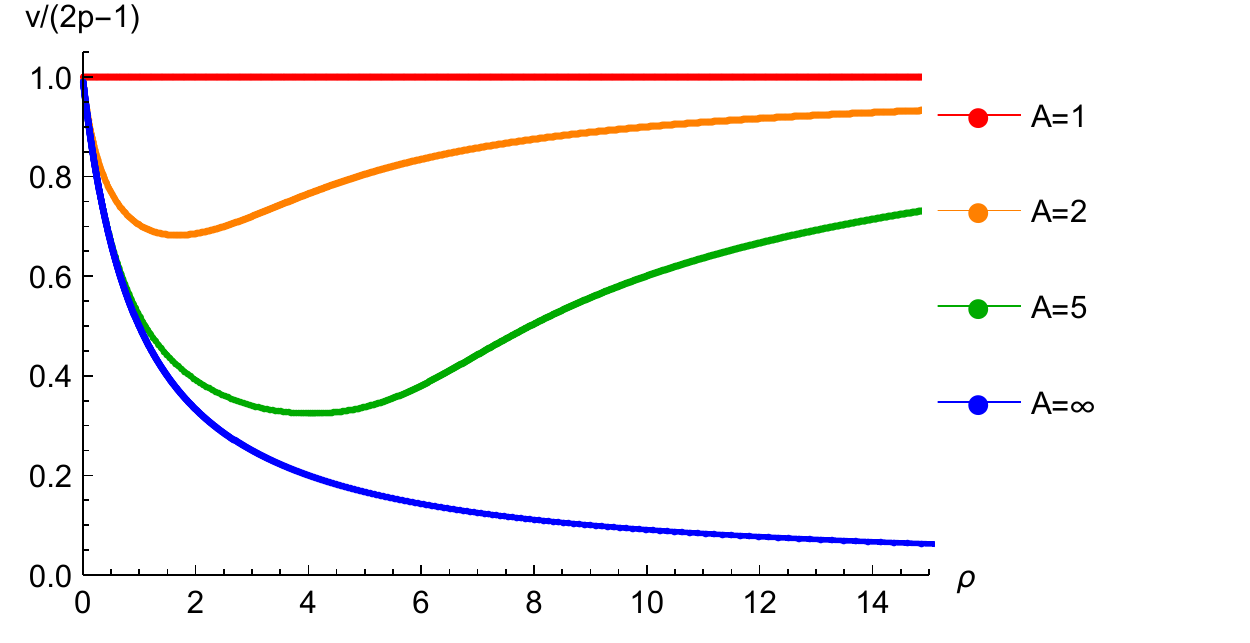}
\vspace{1mm}
\includegraphics[width=7.7cm]{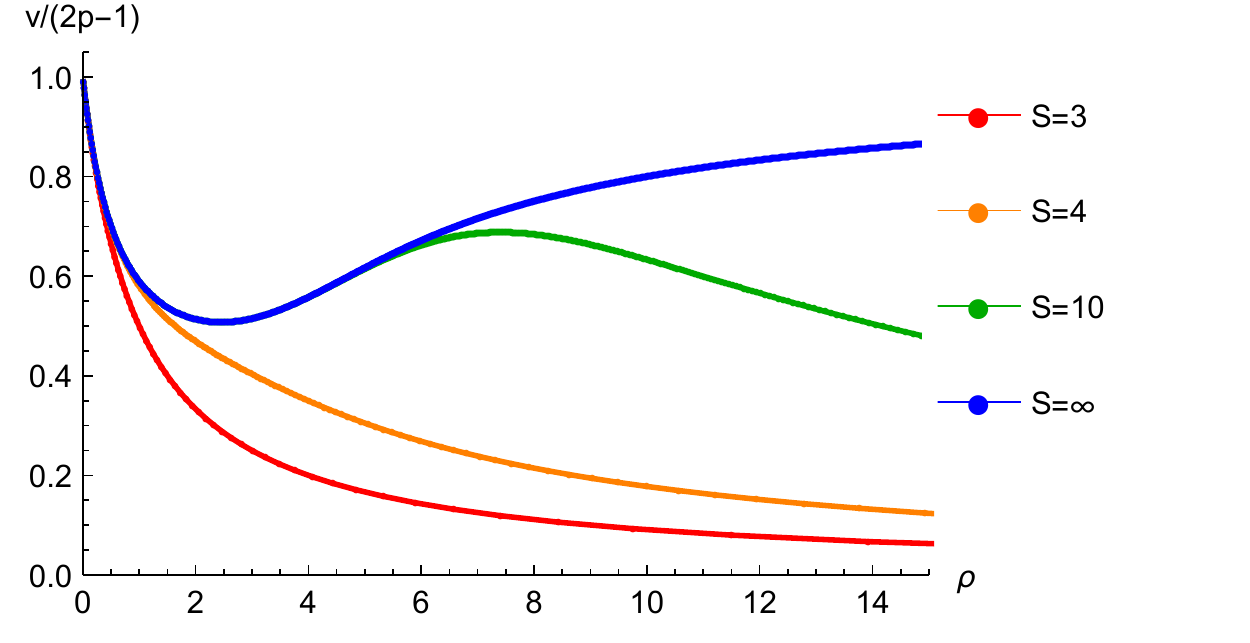}
\hspace{1mm}
\includegraphics[width=7.7cm]{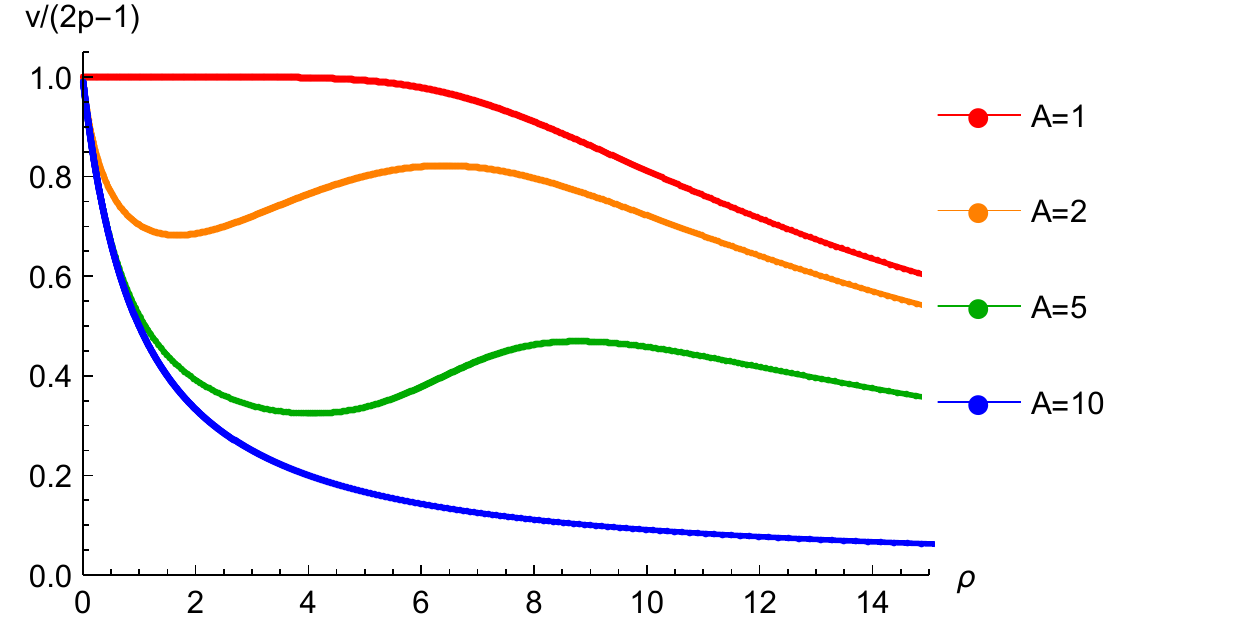}
\caption{\textit{Left panel, top row}: Behavior of the velocity 
$v/(2p-1)$ vs. $\rho$ for $A=1$ and for different 
values of the saturation threshold, 
i.e., $S=1,2,5,\infty$. 
\textit{Right panel, top row}: Behavior of the velocity 
$v/(2p-1)$ vs. $\rho$ for $S=\infty$ and for different 
values of the activation threshold, 
i.e., $A=1,2,5,\infty$. \textit{Left panel, bottom row}: 
Behavior of the velocity 
$v/(2p-1)$ vs. $\rho$ for $A=3$ and for $S=3,4,10,\infty$. 
\textit{Right panel, bottom row}: Behavior of the velocity 
$v/(2p-1)$ vs. $\rho$ for $S=10$ and for $A=1,2,5,10$.}
\label{fig:v}
\end{figure}
It can be proven that the equation governing the evolution of 
the macroscopic local density $\varrho$ is \eqref{diffusion} 
with the 
\textit{macroscopic current} $J(\varrho)$ defined as
\be 
\label{Jns}
J(\varrho)
=
(2p-1)
\nu_{\bar{z}(\rho)}\left[g(\omega_1)\right] 
\ee
where, we recall, the intensity function is defined in \eqref{soglia} and 
the Gibbs measure is defined in \eqref{nu}, see \cite[equation (1.3)]{CR}.

In this out--of--equilibrium regime, the relevant quantity we look at 
is the \textit{velocity}, defined as
$v(\rho)=J(\rho)/\rho$. In particular, it is worth clarifying, here, how the constitutive relation $v$ 
vs.\ $\rho$ is affected by the activation and 
saturation thresholds.
This point may also lead to a more detailed understanding of the 
so--called ``fundamental diagrams'', typically invoked in the context of 
pedestrian flows investigations.

We can now use our results of Section~\ref{s:modello} to compute  
the current. 
First, note that, for any value of the threshold, 
by \eqref{mediaI}, it holds
\begin{equation}
\label{flussom}
J(\rho)=(2p-1)\,\bar{z}(\rho).
\end{equation}
It is not possible to write such an expression explicitly, but for the 
independent particle and simple exclusion--like cases, 
in which cases it is straightforward to derive the well known results 
\begin{equation}
\label{udfipse}
v(\rho)=2p-1
\;\;\;\textrm{ and }\;\;\;
v(\rho)=\frac{2p-1}{1+\rho},
\end{equation}
respectively, where we used the results in Remark~\ref{rem1} and Remark \ref{rem2}.

Figure~\ref{fig:v} shows the behavior of the velocity $v$
as a function of the local density for different values of $A$ and $S$. 
An inspection of the upper left panel of Figure~\ref{fig:v} 
confirms that the velocity 
divided by the bias $(2p-1)$
is equal to unity for the independent particle model, and behaves as $(1+\rho)^{-1}$ in the simple exclusion--like case. 

 Similarly to the case of the diffusion coefficient, we also notice the presence of a 
non--monotonic behavior of $v$ as a function of $\rho$, 
occurring if $A>1$ and
$S\gg A$.
Again, this effect can be ascribed to the peculiar properties of 
the microscopic dynamics, constrained by the two thresholds.

In particular, the right top panel of Figure~\ref{fig:v} 
shows the case $S=\infty$. In absence of limitations due to the exits capacity, 
if no limitation on the communication occurs ($A=1$), 
the typical speed is maximal and it does not depend on the 
local density. 
On the other hand, when $A>1$, the speed decreases until the density exceeds a critical value (depending on the two thresholds), and, after that,
it starts to increase until it attains the ideal maximal value 
at large $\rho$. 
Indeed, if the density is below such a critical value the intensity 
function is equal to one independently on the typical number of on site particles, hence the number of particles that 
leaves a site per unit of time does not depend on the number of particles 
on it. On the other hand, when such a critical value is 
overcome, the density function starts to behave proportionally 
to the number of on site particles and the typical velocity 
starts to increase with the local density.
In the extreme case $A=\infty$, no communication
is possible however large is the density, hence the typical 
velocity is a monotonic decreasing function of $\rho$. 

In the right bottom panel, the case $S=10$ is portrayed: the graphs 
show that as a result of the constraints imposed by the two dynamical thresholds, there exists a local value of the 
density optimizing the typical speed. Such a density has to be large 
enough so that communication is efficient but, also, small enough 
so that the limitation on the escape capacity do not cause an abrupt 
drop of the typical velocity. 

We also run a set of Monte Carlo simulations for a ZRP on a finite lattice equipped with periodic boundary conditions, in order to check the consistency of the results for the velocity $v(\rho)$ obtained above in the hydrodynamic limit.
The dynamics on the finite lattice was performed using the following steps:
\begin{itemize}
\item[(i)] a number $\tau$ is chosen at random with 
exponential distribution of 
parameter $\sum_{x=1}^Lg(\omega_x(t))$, and time is correspondingly updated to  
$t+\tau$;
\item[(ii)]
 a site is chosen at random with probability 
$g(\omega_x(t))/\sum_{x=1}^Lg(\omega_x(t))$;
\item[(iii)] a particle is moved from the selected site to one of its 
nearest neighbors on the right or on the left with probability $p$ or, 
respectively, $1-p$.
\end{itemize}
Starting from an arbitrary initial configuration $\omega_o$ at time $t=0$, the simulation is let then evolve for $n_{tot}\sim10^7$ steps.
The stationary current is then obtained by computing the difference 
between the 
the total number of particles hopping from the site $L$ to the site $1$
and that of particles jumping from $1$ to $L$, 
and dividing, then, the resulting value by the total time. \\
It is worth also remarking that the considered magnitude of $n_{tot}$ was chosen large enough to guarantee the achievement of a stationary value of the current.

\begin{figure}
\centering
\includegraphics[width=7.7cm]{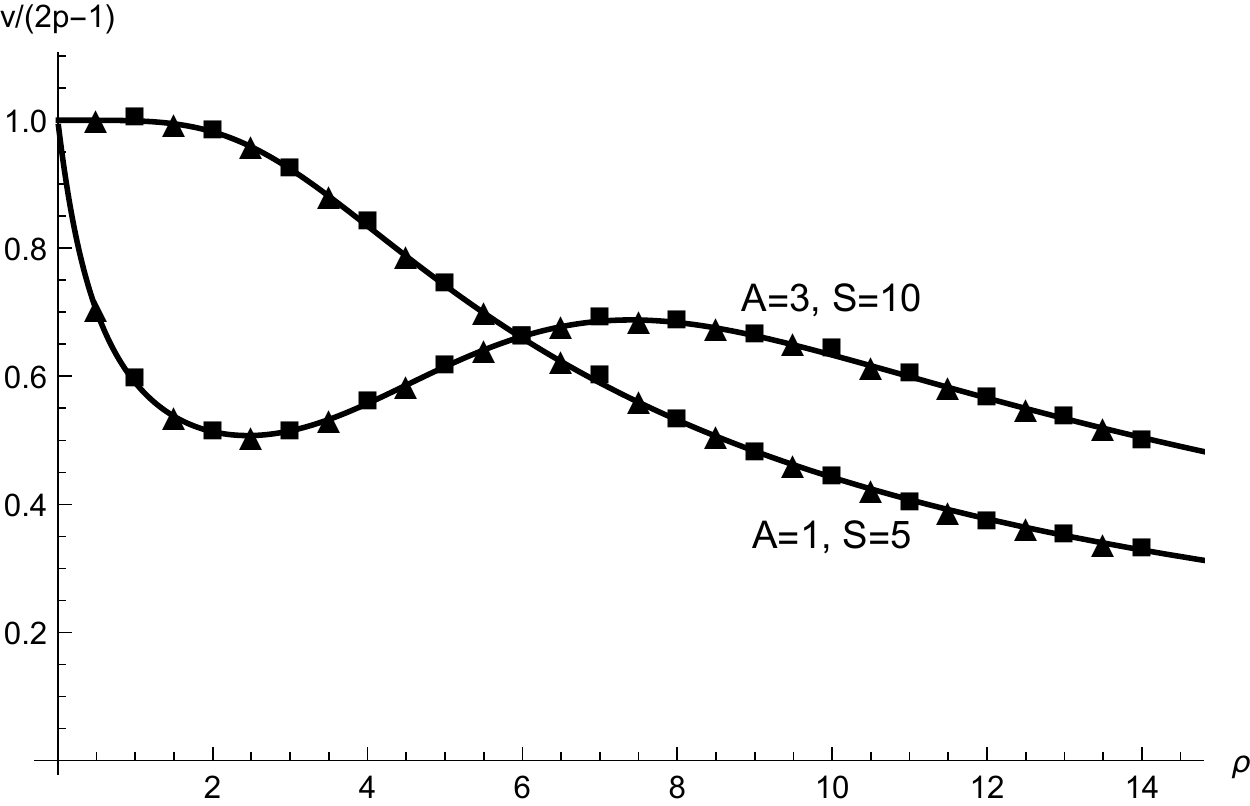}
\hspace{1mm}
\includegraphics[width=7.7cm]{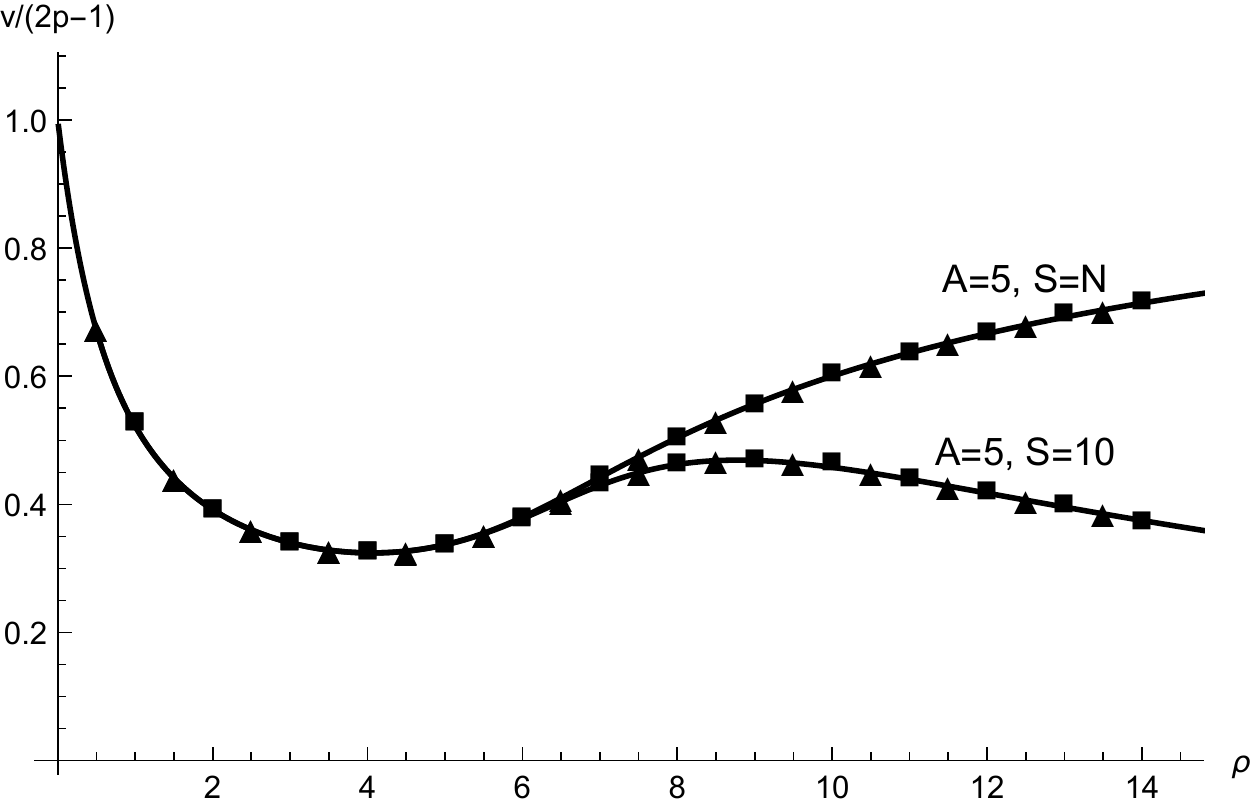}
\caption{Comparison of analytical results for the velocity current in the hydrodynamic limit (cf. Fig. \ref {fig:v}) with Monte Carlo simulations. The squares and the triangles, in the plots, denote the results of simulations obtained with $p=0.6$ and, respectively, $p=0.8$.}
\label{fig:simul}
\end{figure}

In Fig. \ref{fig:simul} the results of the Monte Carlo simulations obtained with $p=0.6$ and $p=0.8$, $L=100$ and for increasing values of $N$, are displayed together with the corresponding curves shown in Fig. \ref{fig:v} and referring to the hydrodynamic limit. The left plot of Fig. \ref{fig:simul} shows two cases, the first corresponding to $A=1$ and $S=5$, whereas the second to $A=3$ and $S=10$. Similarly, the right plot shows two different cases: the first with $A=5$ and $S=10$, while the second with $A=5$ and $S=N$. The plots reveal that our numerical simulations succeed to quantitatively reproduce the predicted behavior of $v(\rho)$ which holds in the hydrodynamic limit, including, in particular, the non-monotonic behavior of the velocity present for finite values of the two thresholds $A$ and $S$.

\section{Possible interpretations of the two thresholds}
\par\noindent
It is worth mentioning that working with two thresholds leads to rich descriptions in terms of modeling. In particular, a double-threshold dynamics is amenable to be interpreted in multiple fashions, viz.
\begin{itemize}
\item[(i)] {\em Porous media interpretation}: Essentially, the bulk porosity estimates how many particles can be accommodated in a cell. This connects to the saturation threshold. The saturation threshold is essentially proportional to the surface porosity, since it is a measure of the exits capacity.  We refer to \cite{Bear} for building a possible closer look on the porous media interpretation. 
\item[(ii)] {\em Mechanical interpretation}: Imagine, for a moment, that the tunnels are equipped with valve-like doors whose opening results from the balance between the pressure inside the cell and an outer pressure exerted by a spring.
A minimal -- structural -- opening of the door, with the spring maintained at rest, corresponds to the presence of an activation threshold. Any further opening of the door is hence achieved by compensating the external pressure of the spring, which is considered to increase proportionally to the displacement of the door, as dictated by the Hooke's law of mechanics. Finally, the maximal opening of the door, in presence of the minimum elongation of the spring, corresponds to the saturation threshold.
See, e.g., \cite{Aiki-rice} for a scenario describing how pressure/temperature-controlled shape--memory alloys facilitate the functioning  of the Japanese rice cooking machine.
\item[(iii)] {\em Psychologico--geometrical interpretation}: 
The activation threshold is a measure of the domain of communication between the individuals and  the level this communication is processed towards a decision on the motion (either on orientation in the dark, or on the chosen speed). Essentially, we imagine that this activation threshold is inversely proportional to {\em the level of trust} (see our interpretations proposed in \cite{CRAS-CCM}). The saturation threshold is then directly proportional to the capacity  of the exit(s). 
\item[(iv)]  {\em A phase transitions perspective}: The assumption here is that pedestrians evacuating the obscure tunnel undergo a first transition of first kind (like the ice-water transition, cf. Landau's classification): from being trapped in the dark tunnel and being free to go in corridors where they can choose their own desired velocity. The parallel can be made a bit more precise by applying the Clapeyron equation in this context to translate difference in temperatures into difference in pressures. The two thresholds can now be seen as the direct counterparts of the accumulated heat content (amount of phonons) needed
to melt the ice (the activation threshold) and the amount of accumulated heat content needed to evaporate water (the saturation threshold). Essentially, we mean here that the dynamics is ``frozen'' for densities below the activation threshold and people ``evaporate'' from the tunnel  for densities of the order of magnitude of the saturation threshold. Remotely related connections to phase transitions supposed to happen 
in social systems are reported, for instance, in \cite{PT1,PT2}. 
\end{itemize}

\section{Discussion}
\par\noindent
\subsection{Multiscale modeling perspectives}
\par\noindent
We considered a one--dimensional ZRP equipped with 
periodic boundary conditions and characterized by symmetric or asymmetric jump probabilities. 
The novelty of our approach stems from introducing the two thresholds $A$ and $S$ affecting the stochastic dynamics, together with their interpretations in terms of communication efficiency and exit capacity. 

From the mathematics viewpoint, the thresholds  can be tuned to control the magnitude of the intensity function, thus allowing one to span a broad variety of zero range dynamics, ranging from the independent particle models to the simple exclusion--like processes. 

We then investigated the hydrodynamic limit of the considered ZRP for different values of the thresholds, and discussed the effect of such dynamical constraints on some macroscopic quantities, e.g. the effective diffusion coefficient, the particle density and the effective outgoing current. We  recovered known results in the limiting scenarios, and also provided explicit formulae for  arbitrary thresholds, provided the activation and saturation thresholds coincide.  Our investigation thus provides a noteworthy bridge between the features of the microscopic stochastic dynamics and some macroscopic observables relevant in the hydrodynamic description of the model, which are also experimentally accessible. Further investigations are needed, next, to extend our results to the even more challenging scenario characterized by the use of non--periodic boundary conditions in the zero range dynamics.

From the pedestrians evacuation viewpoint, we explored the effects of
communication  on the effective transport properties of the crowd of pedestrians. More precisely, we were able to emphasize the effect of two thresholds
on the structure of the effective nonlinear diffusion coefficient. One threshold models pedestrians' communication efficiency in  the dark, while the other one
describes the tunnel capacity.  Essentially, we observe  that  if the evacuees show a maximum trust (leading to a fast communication), they tend to quickly find the exit
and hence the collective action tends to prevent the occurrence of disasters. 
In our context, ``a high activation threshold increases the diffusion coefficient" means that 
``higher trust among pedestrians improves
communication in the dark"
and therefore the exits can be found more easily. The exit capacity is accounted for by the magnitude of the saturation threshold. 
Consequently, a higher saturation threshold leads to an improved capacity of the exists (e.g. larger doors, or more exits \cite{Ronchi}) and, hence, the evacuation rate is correspondingly higher. 

Similarly, in presence of a drift, the fundamental diagrams become 
non--monotonic with respect to the local pedestrian density. We were able to point out that the fundamental diagrams become independent on the local density as soon as the exit capacity is unbounded. 
Interestingly, we were able to detect situations (see, for instance,  
Figure~\ref{fig:v}) in which there are particular pedestrian densities optimizing the 
speed (see e.g. Figure 2b in \cite{Rajat} for real pedestrian traffic cases where this effect has been observed). It appears that such an {\em optimizing density} must be large 
enough so that communication is efficient but, also, small enough 
so that the limitation on the escape capacity do not cause an abrupt 
drop of the typical flow velocity. 

\subsection{Qualitative validation}
\par\noindent
If one wants to make predictions, then models must be calibrated with empirical data.   Designing a crowd experiment to test our
 pedestrians-moving-in-dark model is a challenge from many perspectives (including ethical and practical aspects) 
that we don't undertake here.   As future plan, we wish to adapt our model to make progress toward a quantitative validation for scenarios involving pedestrians moving in regions filled with a dense smoke, with specific reference to the crowd experiments made by the Department of Fire Safety Engineering of 
the Lund  University, Sweden, see e.g. \cite{Ronchi,Ronchi2} and references cited  therein.  In that case, the main target would be to set up a parameter identification procedure at the ZRP level for finding suitable combinations of the thresholds $A$ and $S$ to recover 
typical smoke concentration-dependent speed-density relations (fundamental diagrams). Our simulations based on the current ZRP model with two thresholds give hope in this direction in the sense that, for the drift-dominated dynamics endowed with a full communication among pedestrians (i.e. for $A=1$), we are able to recover for the saturation threshold $S=5$ the same monotonic shape of real pedestrian traffic fundamental diagrams as reported in \cite{Zhang}, e.g. To see this trend, compare Figure~\ref{fig:v} (left panel, top row, $S=5$).  Furthermore, we note in the same Figure that as the density increases  the fundamental diagram tends towards a linear profile regardless of the choice of the threshold $S$. Such situation is considered as standard for pedestrian dynamics, compare for instance  Figure 3.4 on page 33 in \cite{Corbetta} or \cite{SS08}.

\section*{Acknowledgements}
The authors wish to thank Errico Presutti (Gran Sasso Science Institute, 
L'Aquila, Italy), Anna De Masi (University of L'Aquila, Italy), 
and Claudio Landim  (IMPA, Rio de Janeiro, Brazil) for useful discussions. 
ENMC thanks ICMS (TU/e, Eindhoven, The Netherlands) for the very
kind hospitality and for financial support.

\appendix
\section{Some explicit formulas for arbitrary values of the thresholds}
\label{sec:app1}
We provide, here, the general form of the normalization 
constant $C_z$ and the function $\bar{\rho}(z)$ for arbitrary values 
of the thresholds $A$ and $S$.
It holds
\be
C_z=z \left[ \frac{z-z^A}{1-z}+\frac{z^{2+S}}{\Gamma(2+S-A)}\left(\frac{1}{1-z+S-A}+e^z E_{(-1-S+A)}(z)\right)\right]^{-1}\quad , \nonumber
\ee
where
\be
\Gamma(x)=\int_0^\infty s^{x-1}e^{-s}ds \nonumber
\ee
is the Gamma function and
\be
E_{n}(x)=\int_1^\infty \frac{e^{-x s}}{s^n}ds \nonumber
\ee
denotes the generalized exponential integral function \cite{AS}.
Moreover, 
\bea
\bar{\rho}(z)&=&C_z z 
 \left[ \frac{z^2+z^A(1-A(1-z)-2z)}{z(1-z)^2}
 +\frac{z^{1+S}}{\Gamma(2+S-A)}\times\right.\nonumber
\vphantom{\bigg\{_\}}\\
&&\times\left.\left(\frac{z+(A+z)(1-z+S-A)}{(1-z+S-A)^2}+e^z (-1+A+z)E_{(-1-S+A)}(z)\right)\right].\label{den}
\eea
One can then verify that, by taking $A=1$ and $S=\infty$, 
one recovers the expressions for $C_z$ and $ \bar{\rho}(z)$ corresponding 
to the independent particle model (see Remark \ref{rem1}), 
whereas, for $A=S$, one obtains the results pertaining to the simple 
exclusion--like model (see Remark \ref{rem2}).


\end{document}